\documentclass[12pt]{article}
\usepackage{amssymb,amsmath,graphicx}
\usepackage{amsfonts,setspace}
\usepackage{caption,natbib}
\usepackage[caption=false]{subfig}
\usepackage{epsfig,latexsym,graphicx}
\newtheorem{theorem}{Theorem}[section]

\newtheorem{remark}{Remark}[section]

\begin{document}

\title{Robust and Efficient Estimation in the Parametric Cox Regression Model under Random Censoring}

\author{Abhik Ghosh and Ayanendranath Basu\footnote{Corresponding author}\\
Interdisciplinary Statistical Research unit\\
Indian Statistical Institute, Kolkata, India  
\\
{\it abhik.ghosh@isical.ac.in, ayanbasu@isical.ac.in}}
\maketitle

\begin{abstract}
Cox proportional hazard regression model is a popular tool to analyze 
the relationship between a censored lifetime variable with other relevant factors. 
The semi-parametric Cox model is widely used to study different types of data 
arising from applied disciplines like medical science, biology, reliability studies and many more. 
A fully parametric version of the Cox regression model, if properly specified, 
can yield more efficient parameter estimates leading to better insight generation. 
However, the existing maximum likelihood approach of generating inference under 
the fully parametric Cox regression model is highly non-robust against data-contamination
which restricts its practical usage. 
In this paper we develop a robust estimation procedure for the parametric Cox regression model
based on the minimum density power divergence approach. 
The  proposed minimum density power divergence estimator is seen to produce
highly robust estimates under data contamination with only a slight loss in efficiency under pure data.
Further, they are always seen to generate more precise inference 
than the likelihood based estimates under the semi-parametric Cox models or 
their  existing robust versions. We also sketch the derivation of the asymptotic properties of 
the proposed estimator using the martingale approach and 
justify their robustness theoretically through the influence function analysis.  
The practical applicability and usefulness of the proposal are illustrated through simulations 
and a real data example. 
\end{abstract}
\textbf{Keywords:} Minimum Density Power Divergence Estimator; Cox Regression; Parametric Survival Models; 
Robustness; Influence Function; Random Censoring; Counting Process Martingale.

\thispagestyle{empty}

\clearpage
\pagenumbering{arabic} 
\doublespacing
\section{Introduction}\label{SEC:intro}

Randomly censored lifetime data frequently occur in many applications like medical science,
biology, reliability studies, etc., which need to be analyzed properly to 
make correct inference and suitable research conclusions.
The data are often right censored because it is not possible 
to observe the patients or the items under study till their death 
or patients may withdraw during the study period.
Mathematically, if $t_1, \ldots, t_n$ denote the actual life-times of $n$ independent patients (or items) under study,
in reality we only observe $x_i = \min\{t_i, c_i\}$, $i=1, \ldots, n$,
where $c_1, \ldots, c_n$ are their respective censoring times. 
It is generally assumed that $\{t_i\}$, $\{c_i\}$ and $\{x_i\}$ are separately independent and identically distributed (IID) realizations 
of the true lifetime variable $T$, the censoring variable $C$ and 
the observed lifetime variable $X$, respectively, having distribution functions $G_T$, $G_C$, and 
$G_X=1-(1-G_T)(1-G_C)$. 
Generally the censoring information is available, so that 
we know $\delta_i = I(t_i\leq c_i)$ for each $i=1, \ldots, n$,
which can also be thought of as IID realizations of the random variable $\delta=I(T\leq C)$.
Here $I(E)$ denotes the indicator function for the event $E$. 
In the absence of other relevant data, 
one needs to do inference about the true life-time distribution $G_T$ from 
$\left\{(X_i, \delta_i), i=1, \ldots, n\right\}$.
The Kaplan-Meier product limit (KMPL) estimator \citep{Kaplan/Meier:1958} is 
commonly used to non-parametrically estimate $G_T$.
However, a more efficient inference procedure can be derived through the parametric approach 
where one assumes a parametric model for the density $g_T$ of the life-time distribution $G_T$ or 
the corresponding hazard rate $\lambda(t) = g_T(t)/[1-G_T(t)]$. 
These parametric assumptions are often based on previous experiences
(e.g., similar drugs or similar diseases may have been studied in the past);
some commonly used examples are exponential, Weibull, log-normal, log-logistic, gamma, etc.
Such parametric inference procedures under randomly censored data (without any additional covariates)
are well-studied in the literature; 
see \cite{Borgan:1984}, \cite{Andersen/Borgan:1985} and \cite{Hjort:1986a} for the classical maximum likelihood procedures
and \cite{Basu/etc:2006}, \cite{Cherfi:2012}, \cite{Ghosh/etc:2017},  etc.~for more recent robust inference procedures. 
The robust procedures are much more stable in the presence of data contamination;
hence they are more useful in practical applications which are prone to data contamination.

In complex real life scenarios, the life-time variables often depend on several associated factors
which need to be modelled properly for better insight generation. 
For example, life-time of patients in a medical study always depends on patients' age, sex, demography,
and other conditions, along with the treatments, hospital conditions, socio-economic factors, etc. 
In such scenarios, we need to model the life-time variable $T$ given the values of other available covariates,
say $\boldsymbol{Z} \in \mathbb{R}^p$, through an appropriate regression structure. 
Among others, the Cox proportional hazard regression model \citep{Cox:1962}
is widely used in medical and biological applications 
which assumes that the covariate effects on the life-time hazard rate $\lambda(t)$ 
are multiplicative and independent over time $t$.
In particular, assuming that $\boldsymbol{z}_i$ denotes the covariate value of the $i$-th patient,
her hazard rate is modeled by the semi-parametric relation
\begin{equation}
\lambda_i(t) = \lambda(t|\boldsymbol{Z}=\boldsymbol{z}_i) = \lambda_0(t)e^{\boldsymbol{\beta}^T\boldsymbol{z}_i},
~~~i=1, \ldots, n,
\label{EQ:coxReg}
\end{equation}
where $\lambda_0$ is the unknown baseline hazard in the absence of any covariates 
and $\boldsymbol{\beta}\in\mathbb{R}^p$ is the unknown regression coefficients.
These unknowns are estimated based on the observed data 
$\left\{(x_i, \delta_i, \boldsymbol{z}_i) : i=1, \ldots, n, \right\}$
through maximum, partial or conditional likelihood approach; 
see, e.g., \cite{Cox:1972, Cox:1975}, \cite{Andersen/Gill:1982} and \cite{Kalbfleisch/Prentice:1980}.

However, as noted earlier, a properly specified parametric approach 
always yields more efficient inference than any non-parametric or semi-parametric approach.
As discussed by \citet[][ch.~8]{Hosmer/etc:2008},
a fully parametric model has several important advantages including  
(i) greater efficiency, (ii) more meaningful estimates having simple interpretations,
(iii) precise prediction, etc.
Their successful application can be found in 
\cite{Cox/Oakes:1984}, \cite{Crowder/etc:1991}, \cite{Collett:2003}, \cite{Lawless:2003},
\cite{Klein/Moeschberger:2003} among others.  
Further, \cite{Hjort:1992} noted that such a parametric model 
{\it ``would lead to more precise estimation of survival probabilities and related quantities 
	and concurrently contribute to a better understanding of the survival phenomenon under study"
	\cite[][p.~375]{Hjort:1992}.}  
Therefore, for greater efficiency, 
a direct parametric extension of the Cox model (\ref{EQ:coxReg}) can be considered by assuming 
a parametric form for the unknown baseline hazard $\lambda_0(t)$,
i.e., we assume the fully parametric regression structure 
\begin{eqnarray}
\lambda_{i,\boldsymbol{\theta}}(t) = \lambda_{\boldsymbol{\theta}}(t|\boldsymbol{Z}_i) = \lambda(t,\boldsymbol{\gamma})e^{\boldsymbol{\beta}^T\boldsymbol{Z}_i},
~~~i=1, \ldots, n, ~~\boldsymbol{\theta}=(\boldsymbol{\gamma}^T, \boldsymbol{\beta}^T)^T\in\mathbb{R}^{q+p},
\label{EQ:coxReg_parametric}
\end{eqnarray}
where $\lambda(t,\boldsymbol{\gamma})$ is a known parametric function
involving the unknown $\boldsymbol{\gamma}\in\mathbb{R}^q$. 
Common examples of $\lambda(t,\boldsymbol{\gamma})$ could be the hazard rate of 
the standard lifetime distributions like exponential, Weibull, log-normal, etc.
or the piece-wise constant hazard.
Then, the full parameter vector $\boldsymbol{\theta}=(\boldsymbol{\gamma}^T, \boldsymbol{\beta}^T)^T$
can be estimated efficiently based on the observed data 
$\left\{(x_i, \delta_i, \boldsymbol{z}_i) : i=1, \ldots, n \right\}$
through the maximum likelihood approach on which the subsequent inference can be based. 
See \cite{Borgan:1984}, \cite{Hjort:1992} and \cite{Kalbfleisch/Prentice:1980} for detailed properties and applications
of the likelihood based inference under the parametric Cox regression model.

Although asymptotically efficient, a major drawback of this maximum likelihood approach,
used in estimating the Cox regression model, is its high instability under data contamination 
\citep{Reid/Crepeau:1985, Hjort:1992}.
As outliers are not uncommon in modern complex datasets in many applications 
including medical and biological studies, a robust approach under the parametric Cox regression model
would be highly useful to provide the best trade-off between the efficiency and robustness of the deduced inference. 
However, although a few robust alternatives for the semi-parametric Cox regression model (\ref{EQ:coxReg}) exist
\citep{Bednarski:1993,Sasieni:1993a,Sasieni:1993b,Bednarski:2007,Farcomeni/Viviani:2011},
the literature has paid little attention to developing robust estimators 
under their fully parametric version (\ref{EQ:coxReg_parametric}). 
This paper fills this gap in the  literature by developing a robust estimation approach 
for parametric Cox regression in presence of random censoring.

Among several possible approaches to robust inference, we consider the minimum divergence approach 
where the unknown parameter is estimated by minimizing a suitable discrepancy measure
between the observed data and the postulated model.
In particular, we consider the density power divergence (DPD) measure 
originally introduced by \cite{Basu/etc:1998} for complete IID data. 
The DPD, a generalization of the  Kullback-Leibler divergence (KLD), 
is given by
\begin{equation}\label{EQ:dpd}
d_\alpha(g,f) = \displaystyle \int  \left[f^{1+\alpha} - \left(1 + \frac{1}{\alpha}\right)  f^\alpha g + 
\frac{1}{\alpha} g^{1+\alpha}\right]d\mu,~~~\alpha\geq 0,
\end{equation}
for any two densities $g$ and $f$ with respect to some common dominating measure $\mu$. 
As the tuning parameter $\alpha\rightarrow 0$, the DPD measure tends to the KLD measure 
$d_0(g,f) =\int g \log(g/f)d\mu$, whereas $d_1$ coincides with the squared $L_2$ distance.
Note that, the MLE is a minimizer of the KLD measure between the data and the model. 
Therefore the minimum DPD estimator (MDPDE), obtained as the minimizer of the DPD measure between the empirical 
data density and the assumed model density, yields a robust generalization of the MLE;
it coincides with the non-robust MLE at $\alpha=0$ and becomes more robust as $\alpha>0$ increase.
Due to several nice properties \cite[see, e.g.,][]{Basu/etc:2011}, along with its simplicity in construction and computation
(we have a simple unbiased estimating equation as a generalization of the likelihood score equation),
the MDPDE has also been extended successfully to several types of models.
In parametric survival analysis, 
the MDPDE has been developed and successfully applied by \cite{Basu/etc:2006} and \cite{Ghosh/Basu:2017}
for randomly censored variables without covariates and a parametric accelerated failure time model, respectively.

In this paper, we develop the MDPDE for  
the fully parametric Cox regression model (\ref{EQ:coxReg_parametric}) 
based on the randomly censored observations
$\left\{(x_i, \delta_i, \boldsymbol{z}_i) : i=1, \ldots, n \right\}$.
The asymptotic properties of this new MDPDE 
including its consistency and asymptotic normality are derived using the martingale approach.
Its robustness is illustrated theoretically through 
the influence function analysis and numerically through appropriate simulation studies. 
The superior efficiency and robustness of the proposed MDPDE under the fully parametric model (\ref{EQ:coxReg_parametric})
compared to the existing robust estimators under the semi-parametric formulation (\ref{EQ:coxReg}) 
are clearly visible in all our illustrations.
The applicability of the proposed methodology is illustrated with some real data
and the paper ends with a short concluding discussion.

\section{Estimation in Parametric Cox Regression Models}
\label{SEC:MDPDE}

\subsection{The Maximum Likelihood Estimator}\label{SEC:MLE}

For better understanding of the proposed estimator, 
we start by recalling the maximum likelihood estimator (MLE) under 
the fully parametric Cox regression model (\ref{EQ:coxReg_parametric}).
Throughout this paper, we will make the standard assumption that the observed data
$(x_i, \delta_i, \boldsymbol{z}_i)$, for $i=1, \ldots, n,$ are IID realizations of the random variables
$(X, \delta, \boldsymbol{Z})$ having true joint distribution $H$ on $[0,\infty)\times\{0,1\}\times\mathbb{R}^p$,
deduced from $G_T$, $G_C$ and the true distribution $G_Z$ of $\boldsymbol{Z}$.
This IID assumption holds, for example, under random censoring schemes and random covariates. 
For each individual $i$, define the counting process $N_i$ and the at-risk process $Y_i$ as
$
dN_i(s) = I\left\{x_i\in[s, s+ds], \delta_i=1\right\}$, 
$Y_i(s) = I\left\{x_i \geq s\right\},
$
so that the process $M_i(t) = N_i(t) - \int_0^tY_i(s)\lambda_i(s)ds$ is a martingale.
When the data are IID, the sequence $\left\{(N_i, Y_i, M_i) : i=1, \ldots, n\right\}$ also becomes IID and 
we can apply martingale limit theorems under standard regularity conditions \cite[see, e.g.,][]{Billingsley:1968}.
Note that, $M_i$ involves the true hazard rate $\lambda_i$ of $i$-th individual 
and not the model hazard rate $\lambda_{i,\boldsymbol{\theta}}$.
We wish to model this true conditional hazard $\lambda_i$ by the 
parametric Cox regression model (\ref{EQ:coxReg_parametric}).
However, as in usual practice, we will not assume any model for the covariate distribution $G_Z$
and work with the conditional densities given the covariate values.

Now, under the model hazard rate  given by (\ref{EQ:coxReg_parametric}),
the model survival function of $T$ given $\boldsymbol{Z}=\boldsymbol{z_i}$ has the form
$S_{i,\boldsymbol{\theta}}(t) = S_{\boldsymbol{\theta}}(t|\boldsymbol{Z}=\boldsymbol{z}_i) 
= \exp\left[-\Lambda_{\boldsymbol{\gamma}}(t)e^{\boldsymbol{\beta}^T\boldsymbol{Z}_i}\right]$,
where $\Lambda_{\boldsymbol{\gamma}}(t)= \int_0^t \lambda(s, \boldsymbol{\gamma})ds$ is 
the (model) cumulative baseline hazard. 
Therefore, for each $i$,  the model (partial) likelihood of the observed data-point $(x_i, \delta_i)$ 
given the covariate value $\boldsymbol{Z} = \boldsymbol{z}_i$ has the form
\citep{Andersen/etc:1992}
\begin{eqnarray}
f_{i,\boldsymbol{\theta}}(x,\delta)=f_{\boldsymbol{\theta}}(x,\delta|\boldsymbol{Z}=\boldsymbol{z}_i)
=\left[\lambda(x,\boldsymbol{\gamma})e^{\boldsymbol{\beta}^T\boldsymbol{z}_i}Y_i(x)\right]^\delta
\exp\left[-\Lambda_{\boldsymbol{\gamma}}(x)e^{\boldsymbol{\beta}^T\boldsymbol{z}_i}\right],
\label{EQ:coxModel_den}
\end{eqnarray}
where the parameter of interest is $\boldsymbol{\theta}=(\boldsymbol{\gamma}^T, \boldsymbol{\beta}^T)^T$.
Note that, in this set-up with given covariate values, the observations $(X_i, \delta_i)$
are independent but non-homogeneous having true densities 
$g_i(x,\delta) = g(x,\delta|\boldsymbol{Z}=\boldsymbol{z}_i)$,
which we wish to model by the density $f_{i,\boldsymbol{\theta}}$ in (\ref{EQ:coxModel_den}).

The MLE of $\boldsymbol{\theta}$ is defined as the maximizer of
the likelihood function $L_n = \prod_{i=1}^n f_{i,\boldsymbol{\theta}}(x_i,\delta_i)$,
or equivalently as the maximizer of the log-likelihood function
$\frac{1}{n}\log L_n(\boldsymbol{\theta}) $=$ \frac{1}{n} \sum\limits_{i=1}^n  \log f_{i,\boldsymbol{\theta}}(x_i,\delta_i)
$ = Constant $- \frac{1}{n} \sum_{i=1}^n d_0(\widehat{g}_i,f_{i,\boldsymbol{\theta}}),
$
where $\widehat{g}_i$ is the empirical estimate of $g_i$ and $d_0$ is the KLD measure.
Under standard differentiability assumptions, the MLE can be obtained as a solution to the score equation
$\frac{1}{n} \sum_{i=1}^n \boldsymbol{u}_{i,\boldsymbol{\theta}}(x_i,\delta_i) =\boldsymbol{0}_{p+q}$,
where $\boldsymbol{0}_{p+q}$ is the zero vector of length $(p+q)$ and the $i$-th score function 
$\boldsymbol{u}_{i,\boldsymbol{\theta}}(x,\delta) 
= \left( \boldsymbol{u}_{i,\boldsymbol{\theta}}^{(1)}(x,\delta)^T, \boldsymbol{u}_{i,\boldsymbol{\theta}}^{(2)}(x,\delta)^T\right)^T 
= \left(\frac{\partial}{\partial\boldsymbol{\gamma}^T}\log f_{i,\boldsymbol{\theta}}(x,\delta), 
\frac{\partial}{\partial\boldsymbol{\beta}^T}\log f_{i,\boldsymbol{\theta}}(x,\delta) \right)$
given the covariate value $\boldsymbol{Z}=\boldsymbol{z}_i$ has the form
\begin{eqnarray}
\boldsymbol{u}_{i,\boldsymbol{\theta}}^{(1)}(x,\delta) 
&=& \left\{\psi_{\boldsymbol{\gamma}}(x)-\Psi_{\boldsymbol{\gamma}}(x)e^{\boldsymbol{\beta}^T\boldsymbol{z}_i}\right\}
I(\delta=1)-\Psi_{\boldsymbol{\gamma}}(x)e^{\boldsymbol{\beta}^T\boldsymbol{z}_i}I(\delta=0),\\
\boldsymbol{u}_{i,\boldsymbol{\theta}}^{(2)}(x,\delta) 
&=& \boldsymbol{z}_i\left\{1-\Lambda_{\boldsymbol{\gamma}}(x)e^{\boldsymbol{\beta}^T\boldsymbol{z}_i}\right\}I(\delta=1)
-\boldsymbol{z}_i\Lambda_{\boldsymbol{\gamma}}(x)e^{\boldsymbol{\beta}^T\boldsymbol{z}_i}I(\delta=0),
\end{eqnarray}
with $\psi_{\boldsymbol{\gamma}}(x)=\frac{\partial}{\partial\boldsymbol{\gamma}}\log \lambda(x,\boldsymbol{\gamma})$
and $\Psi_{\boldsymbol{\gamma}}(x)=\int_0^x\frac{\partial}{\partial\boldsymbol{\gamma}}\lambda(s,\boldsymbol{\gamma})ds 
= \int_0^x\psi_{\boldsymbol{\gamma}}(s)\lambda(s,\boldsymbol{\gamma})ds$.

The asymptotic distribution of this MLE at the model and outside the model can be found in 
\cite{Andersen/Gill:1982}, \cite{Borgan:1984}, \cite{Andersen/Borgan:1985}, 
\cite{Lin/Wei:1989} and \cite{Hjort:1992}. The main idea is to write down the objective function 
and the estimating equations in terms of the counting processes $N_i$ and $Y_i$ 
as given by 
\begin{eqnarray}
\frac{1}{n}\log L_n(\boldsymbol{\theta}) &=& 
\frac{1}{n} \sum_{i=1}^n  \int_0^T \left[\log\lambda_{i,\boldsymbol{\theta}}(s) I(\delta_i=1) 
+\log S_{i,\boldsymbol{\theta}}(s)\right]I\left(x_i\in[s, s+ds]\right)ds,
\nonumber\\
&=& \frac{1}{n} \sum_{i=1}^n  \int_0^T \left[\left(\log\lambda(s,\gamma)+ \boldsymbol{\beta}^T\boldsymbol{z}_i\right) dN_i(s) 
- Y_i(s)\lambda(s,\gamma)e^{\boldsymbol{\beta}^T\boldsymbol{z}_i}ds\right],
\label{EQ:coxmodel_loglikM}
\\
\boldsymbol{u}_{i,\boldsymbol{\theta}}^{(1)}(x_i,\delta_i)  
&=& \int_0^T \psi_{\boldsymbol{\gamma}}(s) 
\left[dN_i(s)  - Y_i(s)\lambda(s,\gamma)e^{\boldsymbol{\beta}^T\boldsymbol{z}_i}ds\right],\nonumber\\
\boldsymbol{u}_{i,\boldsymbol{\theta}}^{(2)}(x_i,\delta_i)  
&=&  \int_0^T \boldsymbol{z}_i 
\left[dN_i(s)  - Y_i(s)\lambda(s,\gamma)e^{\boldsymbol{\beta}^T\boldsymbol{z}_i}ds\right],\nonumber
\end{eqnarray}
where it is assumed that the process is observed in the time-interval $[0, T]$
and then use appropriate limit theorems for these processes and the associated martingale $M_i$ 
\citep{Billingsley:1961, Billingsley:1968, Gill:1984}.
However, the major drawback of this MLE is that it has an unbounded influence function \citep{Hampel/etc:1986} 
as illustrated by \cite{Reid/Crepeau:1985}, \cite{Lin/Wei:1989} and \cite{Hjort:1992} among others,
which implies its non-robust nature against outliers. Any inference based on this MLE is then also non-robust. 

\subsection{The Proposed Minimum DPD Estimator}

We are now in a position to define the MDPDE for the parametric Cox regression model (\ref{EQ:coxReg_parametric}). 
Since the observed data-points $(x_i, \delta_i)$, given their respective covariate values $\boldsymbol{z}_i$, 
are non-homogeneous under (\ref{EQ:coxReg_parametric}), 
we cannot directly apply the \cite{Basu/etc:1998}  definition of MDPDE for IID data.
An extended definition of the MDPDE under the general non-homogeneous data (without censoring) has been 
 developed by \cite{Ghosh/Basu:2013}, who obtain the MDPDE as the minimizer
of the average of the DPD measures between different estimated true densities and 
the respective model densities. We follow this extended approach to define the MDPDE under the 
parametric Cox model as the minimizer of $\frac{1}{n} \sum_{i=1}^n d_\alpha(\widehat{g}_i,f_{i,\boldsymbol{\theta}})$
with respect to  $\boldsymbol{\theta}=(\boldsymbol{\gamma}^T, \boldsymbol{\beta}^T)^T$ 
for any fixed $\alpha\geq 0$. 

This objective function given by the average DPD measure also has another 
justification as a generalization of the MLE objective function.
Note that, 
the MLE  is also the minimizer of the average (over the unknown covariate distributions)
KLD measures between conditional empirical and model model densities.
Since the DPD  is a generalization of the KLD,
it is intuitive to construct a generalization of the MLE 
by minimizing  the average DPD measure 
$\frac{1}{n} \sum_{i=1}^n d_\alpha(\widehat{g}_i,f_{i,\boldsymbol{\theta}})$
with respect to the parameter of interest. 
Also, whenever the covariates are stochastic, this quantity 
gives (in probability limit) the empirical estimate of the expected population divergence
$E_{G_Z}\left[d_\alpha(g(\cdot|\mathbf{Z}),f_{\boldsymbol{\theta}}(\cdot|\boldsymbol{Z})\right]$;
this is  again an intuitive quantity to minimize for estimating 
the model parameter $\boldsymbol{\theta}$.

Now, using the form of DPD measure given in (\ref{EQ:dpd}),
we note that the third term has no contribution in the minimization with respect to $\boldsymbol{\theta}$
and hence the MDPDE can equivalently be obtained by minimizing the 
simpler objective function 
\begin{eqnarray}
H_{n,\alpha}(\boldsymbol{\theta}) &=& 
\frac{1}{n} \sum_{i=1}^n 
\left[\int f_{i,\boldsymbol{\theta}}^{1+\alpha}  
- \frac{1+\alpha}{\alpha} \lambda_{i,\boldsymbol{\theta}}(x_i)^{\alpha\delta_i} S_{i,\boldsymbol{\theta}}(x_i)^\alpha\right] 
\label{EQ:MDPDE_ObjFunc}
\end{eqnarray}
It is straightforward  to verify that $H_{n,\alpha}(\boldsymbol{\theta})+\frac{1}{\alpha}$ tends to $-\frac{1}{n}\log L_n(\boldsymbol{\theta})$,
as $\alpha\rightarrow 0$; thus the  MDPDE at $\alpha=0$ is nothing but the usual MLE. 
Under standard differentiability assumptions, 
we can alternatively  obtain the MDPDE by solving the system of estimating equations
$ 
\left( \boldsymbol{u}_{n}^{(1, \alpha)}(\boldsymbol{\theta})^T, \boldsymbol{u}_{n}^{(2, \alpha)}(\boldsymbol{\theta})^T\right)^T
:= -\frac{1}{1+\alpha }
\left(\frac{\partial}{\partial\boldsymbol{\gamma}^T}H_{n,\alpha}(\boldsymbol{\theta}), 
\frac{\partial}{\partial\boldsymbol{\beta}^T}H_{n,\alpha}(\boldsymbol{\theta}) \right)^T = \boldsymbol{0}_{p+q}.
$
Some algebra based on (\ref{EQ:MDPDE_ObjFunc})
lead to the simpler form of these estimating equations as given by 
\begin{eqnarray}
\boldsymbol{u}_{n}^{(1, \alpha)}(\boldsymbol{\theta}) &=& \frac{1}{n} \sum_{i=1}^n 
\left[\left\{ \psi_{\boldsymbol{\gamma}}(x_i)\lambda_{i,\boldsymbol{\theta}}(x_i)^\alpha S_{i,\boldsymbol{\theta}}(x_i)^\alpha
-\left(\lambda_{i,\boldsymbol{\theta}}(x)^\alpha-1\right) \Psi_{\boldsymbol{\gamma}}(x_i)
e^{\boldsymbol{\beta}^T\boldsymbol{z}_i}S_{i,\boldsymbol{\theta}}(x_i)^\alpha\right\}I(\delta_i=1)\right.
\nonumber\\
&& ~~~~~~~~~~~ -\left. \Psi_{\boldsymbol{\gamma}}(x_i)e^{\boldsymbol{\beta}^T\boldsymbol{z}_i}
S_{i,\boldsymbol{\theta}}(x_i)^\alpha -\xi_{i}^{(1,\alpha)}(\boldsymbol{\theta})\right] = \boldsymbol{0}_q,
\label{EQ:MDPDE_EstEqn1}\\
\boldsymbol{u}_{n}^{(2, \alpha)}(\boldsymbol{\theta}) &=& \frac{1}{n} \sum_{i=1}^n 
\left[\left\{ \lambda_{i,\boldsymbol{\theta}}(x_i)^\alpha S_{i,\boldsymbol{\theta}}(x_i)^\alpha
-\left(\lambda_{i,\boldsymbol{\theta}}(x)^\alpha-1\right)
\Lambda_{\boldsymbol{\gamma}}(x_i)e^{\boldsymbol{\beta}^T\boldsymbol{z}_i}S_{i,\boldsymbol{\theta}}(x_i)^\alpha\right\}
\boldsymbol{z}_iI(\delta_i=1)\right.
\nonumber\\
&& ~~~~~~~~~~~ 
-\left. \boldsymbol{z}_i\Lambda_{\boldsymbol{\gamma}}(x_i)e^{\boldsymbol{\beta}^T\boldsymbol{z}_i}S_{i,\boldsymbol{\theta}}(x_i)^\alpha 
-\xi_{i}^{(2,\alpha)}(\boldsymbol{\theta})\right] =\boldsymbol{0}_p,
\label{EQ:MDPDE_EstEqn2}
\end{eqnarray}
where 
$\xi_{i}^{(j,\alpha)}(\boldsymbol{\theta}) = \int \boldsymbol{u}_{i,\boldsymbol{\theta}} f_{i,\boldsymbol{\theta}}^{1+\alpha}$, for $j=1,2$.
Again, it follows that $\xi_{i}^{(j,0)}(\boldsymbol{\theta}) =\boldsymbol{0}$ 
and $\boldsymbol{u}_{n}^{(j, 0)}(\boldsymbol{\theta}) 
= \frac{1}{n} \sum\limits_{i=1}^n \boldsymbol{u}_{i,\boldsymbol{\theta}}^{(j)} (x_i, \delta_i)$ for each $j=1, 2$.
Thus, the above MDPDE estimating equations (\ref{EQ:MDPDE_EstEqn1})--(\ref{EQ:MDPDE_EstEqn2})
are indeed a generalization of the MLE score equations in order to achieve greater robustness. 
They are also unbiased at the model distribution as shown in the next section.

\section{Asymptotic Properties of the MDPDE}\label{SEC:asymp}

In order to derive the asymptotic properties of the proposed MDPDEs,
we adopt the martingale approach of \cite{Andersen/Gill:1982}. 
For simplicity in presentation, we here discuss the main asymptotic properties of our MDPDE in a simpler language
with easier assumptions; further basic sufficient conditions for our assumptions can be obtained 
along the lines of \cite{Borgan:1984}, \cite{Andersen/Borgan:1985}, \cite{Hjort:1986a} or \cite{Andersen/etc:1992}. 
However, we develop these asymptotic results under a general class of underlying true distributions
beyond only the model family, which is defined through the following assumption.

\noindent\textbf{Assumption (A):}  
The true hazard rate, given covariate value $\boldsymbol{Z}=\boldsymbol{z}_i$, is of the form 
$\lambda_i(s) = \lambda_0(s)h_0(\boldsymbol{z}_i)$ for some positive functions $\lambda_0$ and $h_0$.  
Denote $\Lambda_0(t) =\int_0^t\lambda_0(s)ds$.

Let us rewrite the left-hand sides of the MDPDE estimating equations 
(\ref{EQ:MDPDE_EstEqn1})--(\ref{EQ:MDPDE_EstEqn2}) 
in terms of the processes $(N_i, Y_i, M_i)$ as 
$
\boldsymbol{u}_{n}^{(j, \alpha)}(\boldsymbol{\theta})
= \frac{1}{n} \sum_{i=1}^n \boldsymbol{u}_{n, i}^{(j, \alpha)}(\boldsymbol{\theta}),
$ 
for each $j=1,2$,
where 

\begin{eqnarray}
\boldsymbol{u}_{n, i}^{(1, \alpha)}(\boldsymbol{\theta}) &=& 
 \left[\int_0^T \left\{ \psi_{\boldsymbol{\gamma}}(s)\lambda(s,\boldsymbol{\gamma})^\alpha 
e^{\alpha\boldsymbol{\beta}^T\boldsymbol{z}_i}S_{i,\boldsymbol{\theta}}(s)^\alpha 
-\left(\lambda(s,\boldsymbol{\gamma})^\alpha 
e^{\alpha\boldsymbol{\beta}^T\boldsymbol{z}_i}-1\right)\Psi_{\boldsymbol{\gamma}}(s)e^{\boldsymbol{\beta}^T\boldsymbol{z}_i}
S_{i,\boldsymbol{\theta}}(s)^\alpha\right\}dN_i(s)\right.
\nonumber\\
&& ~~~~~~~~~~~ -\left. \int_0^T \Psi_{\boldsymbol{\gamma}}(s)e^{\boldsymbol{\beta}^T\boldsymbol{z}_i}S_{i,\boldsymbol{\theta}}(s)^\alpha 
I\left(x_i\in[s, s+ds]\right)-\xi_{i}^{(1,\alpha)}(\boldsymbol{\theta})\right],
\label{EQ:MDPDE_EstEqn1M}\\
\boldsymbol{u}_{n,i}^{(2, \alpha)}(\boldsymbol{\theta}) &=& 
\left[\int_0^T\left\{ \lambda(s,\boldsymbol{\gamma})^\alpha 
e^{\alpha\boldsymbol{\beta}^T\boldsymbol{z}_i}S_{i,\boldsymbol{\theta}}(s)^\alpha-\left(\lambda(s,\boldsymbol{\gamma})^\alpha 
e^{\alpha\boldsymbol{\beta}^T\boldsymbol{z}_i}-1\right)\Lambda_{\boldsymbol{\gamma}}(s)
e^{\boldsymbol{\beta}^T\boldsymbol{z}_i}S_{i,\boldsymbol{\theta}}(s)^\alpha\right\}
\boldsymbol{z}_idN_i(s)\right.
\nonumber\\
&& ~~~~~~~~~~~ 
-\left. \int_0^T\boldsymbol{z}_i\Lambda_{\boldsymbol{\gamma}}(s)e^{\boldsymbol{\beta}^T\boldsymbol{z}_i}S_{i,\boldsymbol{\theta}}(s)^\alpha 
I\left(x_i\in[s, s+ds]\right)-\xi_{i}^{(2,\alpha)}(\boldsymbol{\theta})\right].
\label{EQ:MDPDE_EstEqn2M}
\end{eqnarray}
In order to study their limits, we need some additional notations; 
for each $j=0, 1$ and $\alpha_1, \alpha_2 \geq 0$, let us  denote
$dG_{n,\alpha_1,\alpha_2}^{(j)}(s) = \frac{1}{n} \sum\limits_{i=1}^n (\boldsymbol{z}_i)^j
e^{\alpha_1\boldsymbol{\beta}^T\boldsymbol{z}_i}S_{i,\boldsymbol{\theta}}(s)^{\alpha_2}dN_i(s)$,
$dH_{n,\alpha_1,\alpha_2}^{(j)}(s) = \frac{1}{n} \sum\limits_{i=1}^n (\boldsymbol{z}_i)^j 
e^{\alpha_1\boldsymbol{\beta}^T\boldsymbol{z}_i}S_{i,\boldsymbol{\theta}}(s)^{\alpha_2}I\left(x_i\in[s, s+ds]\right)$,
$Q_{n,\alpha_1,\alpha_2}^{(j)}(s) = \frac{1}{n} \sum\limits_{i=1}^n (\boldsymbol{z}_i)^j 
e^{(\alpha_1+1)\boldsymbol{\beta}^T\boldsymbol{z}_i}S_{i,\boldsymbol{\theta}}(s)^{\alpha_2}Y_i(s)$,
$r_{\alpha_1,\alpha_2}^{(j)}(s) = E\left[ (\boldsymbol{Z})^jI(X\geq s) e^{\alpha_1\boldsymbol{\beta}^T\boldsymbol{Z}}
S_{\boldsymbol{\theta}}(s|\boldsymbol{Z})^{\alpha_2}h_0(\boldsymbol{Z})\right]$,
$q_{\alpha_1,\alpha_2}^{(j)}(s) = E\left[ (\boldsymbol{Z})^jI(X\geq s) e^{(\alpha_1+1)\boldsymbol{\beta}^T\boldsymbol{Z}}S_{\boldsymbol{\theta}}(s|\boldsymbol{Z})^{\alpha_2}\right]$.
In terms of these quantities, we can rewrite $\boldsymbol{u}_{n}^{(j, \alpha)}(\boldsymbol{\theta})$,
$j=1,2$, as 
\begin{eqnarray}
\boldsymbol{u}_{n}^{(1, \alpha)}(\boldsymbol{\theta}) &=& 
\left[\int_0^T \left\{ \psi_{\boldsymbol{\gamma}}(s)\lambda(s,\boldsymbol{\gamma})^\alpha dG_{n,\alpha,\alpha}^{(0)}(s)
-\lambda(s,\boldsymbol{\gamma})^\alpha \Psi_{\boldsymbol{\gamma}}(s)dG_{n,\alpha+1,\alpha}^{(0)}(s)
+ \Psi_{\boldsymbol{\gamma}}(s)dG_{n,1,\alpha}^{(0)}(s)\right\}\right.
\nonumber\\
&& ~~~~~~~~~~~ -\left. \int_0^T\Psi_{\boldsymbol{\gamma}}(s)dH_{n,1,\alpha}^{(0)}(s)
-\xi_{i}^{(1,\alpha)}(\boldsymbol{\theta})\right],
\label{EQ:MDPDE_EstEqn1Ms}\\
\boldsymbol{u}_{n}^{(2, \alpha)}(\boldsymbol{\theta}) &=& 
\left[\int_0^T\left\{ \lambda(s,\boldsymbol{\gamma})^\alpha dG_{n,\alpha,\alpha}^{(1)}(s)
-\lambda(s,\boldsymbol{\gamma})^\alpha \Lambda_{\boldsymbol{\gamma}}(s)dG_{n,\alpha+1,\alpha}^{(1)}(s)
+ \Lambda_{\boldsymbol{\gamma}}(s)dG_{n,1,\alpha}^{(0)}(s)\right\}\right.
\nonumber\\
&& ~~~~~~~~~~~ 
-\left. \int_0^T\Lambda_{\boldsymbol{\gamma}}(s)dH_{n,1,\alpha}^{(1)}(s)-\xi_{i}^{(2,\alpha)}(\boldsymbol{\theta})\right].
\label{EQ:MDPDE_EstEqn2Ms}
\end{eqnarray}
Now, let us assume the following limiting results, along with Assumption (A).

\noindent\textbf{Assumption (B):}  
As $n\rightarrow\infty$, 
$dG_{n,\alpha_1,\alpha_2}^{(j)}(s) \displaystyle\mathop{\rightarrow}^\mathcal{P} r_{\alpha_1,\alpha_2}^{(j)}(s)\lambda_0(s)ds$,
$Q_{n,\alpha_1,\alpha_2}^{(j)}(s) \displaystyle\mathop{\rightarrow}^\mathcal{P} q_{\alpha_1,\alpha_2}^{(j)}(s),
$
and 
$
dH_{n,\alpha_1,\alpha_2}^{(j)}(s) \displaystyle\mathop{\rightarrow}^\mathcal{P}   r_{\alpha_1,\alpha_2}^{(j)}(s)\lambda_0(s)ds
+ q_{\alpha_1-1,\alpha_2}^{(j)}(s)ds.\nonumber
$

Note that Assumption (B) holds under mild boundedness conditions 
on the parametric baseline hazard and the covariate values
by using the limit theorems for empirical processes \citep{Billingsley:1968}.
Further, under Assumptions (A) and (B), 
the quantities $\boldsymbol{u}_{n}^{(j, \alpha)}(\boldsymbol{\theta})$
converges in probability to $\boldsymbol{u}_0^{(j, \alpha)}(\boldsymbol{\theta})$, for each $j=1,2$, respectively, where
\begin{eqnarray}
\boldsymbol{u}_{0}^{(1, \alpha)}(\boldsymbol{\theta}) &=& 
\left[\int_0^T \left\{ \psi_{\boldsymbol{\gamma}}(s)\lambda(s,\boldsymbol{\gamma})^\alpha r_{\alpha,\alpha}^{(0)}(s)
-\lambda(s,\boldsymbol{\gamma})^\alpha \Psi_{\boldsymbol{\gamma}}(s)r_{\alpha+1,\alpha}^{(0)}(s)
\right\}\lambda_0(s)ds\right.
\nonumber\\
&& ~~~~~~~~~~~ -\left. \int_0^T\Psi_{\boldsymbol{\gamma}}(s)q_{0,\alpha}^{(0)}(s)ds
-\xi_{0}^{(1,\alpha)}(\boldsymbol{\theta})\right],
\label{EQ:MDPDE_EstEqn1Msl}\\
\boldsymbol{u}_{0}^{(2, \alpha)}(\boldsymbol{\theta}) &=& 
\left[\int_0^T\left\{ \lambda(s,\boldsymbol{\gamma})^\alpha r_{\alpha,\alpha}^{(1)}(s)
-\lambda(s,\boldsymbol{\gamma})^\alpha \Lambda_{\boldsymbol{\gamma}}(s)r_{\alpha+1,\alpha}^{(1)}(s)
\right\}\lambda_0(s)ds\right.
\nonumber\\
&& ~~~~~~~~~~~ 
-\left. \int_0^T\Lambda_{\boldsymbol{\gamma}}(s)q_{0,\alpha}^{(1)}(s)ds-\xi_{0}^{(2,\alpha)}(\boldsymbol{\theta})\right],
\label{EQ:MDPDE_EstEqn2Msl}
\end{eqnarray}
and $\xi_{0}^{(j,\alpha)}(\boldsymbol{\theta}) 
= \lim\limits_{n\rightarrow\infty}\frac{1}{n}\sum_{i=1}^{n}\xi_{i}^{(j,\alpha)}(\boldsymbol{\theta})
=E\xi_{i}^{(j,\alpha)}(\boldsymbol{\theta})$ for $j=1,2$.
Further calculations yield
\begin{eqnarray}
\xi_{0}^{(1,\alpha)}(\boldsymbol{\theta}) &=& \int_0^T \left\{ 
\psi_{\boldsymbol{\gamma}}(s)\lambda(s,\boldsymbol{\gamma})^{\alpha+1} \widetilde{q}_{\alpha,\alpha}^{(0)}(s)
-\lambda(s,\boldsymbol{\gamma})^{\alpha+1} \Psi_{\boldsymbol{\gamma}}(s)\widetilde{q}_{\alpha+1,\alpha}^{(0)}(s)
-\Psi_{\boldsymbol{\gamma}}(s)\widetilde{q}_{0,\alpha}^{(0)}(s)\right\}ds,
\nonumber\label{EQ:xi01}\\
\xi_{0}^{(2,\alpha)}(\boldsymbol{\theta}) &=& \int_0^T\left\{ 
\lambda(s,\boldsymbol{\gamma})^{\alpha+1} \widetilde{q}_{\alpha,\alpha}^{(1)}(s)
-\lambda(s,\boldsymbol{\gamma})^{\alpha+1} \Lambda_{\boldsymbol{\gamma}}(s)\widetilde{q}_{\alpha+1,\alpha}^{(1)}(s)
-\Lambda_{\boldsymbol{\gamma}}(s)\widetilde{q}_{0,\alpha}^{(1)}(s)\right\}ds,
\nonumber\label{EQ:xi02}
\end{eqnarray}
where 
$
\widetilde{q}_{\alpha_1,\alpha_2}^{(j)}(s) 
= E\left[ (\boldsymbol{Z})^je^{(\alpha_1+1)\boldsymbol{\beta}^T\boldsymbol{Z}}
S_{\boldsymbol{\theta}}(s|\boldsymbol{Z})^{\alpha_2+1}\right].
$
Substituting in Eqs.~(\ref{EQ:MDPDE_EstEqn1Msl})--(\ref{EQ:MDPDE_EstEqn2Msl}), 
we get 

\begin{eqnarray}
&&\boldsymbol{u}_{0}^{(1, \alpha)}(\boldsymbol{\theta}) = 
\int_0^T \left[\psi_{\boldsymbol{\gamma}}(s)\lambda(s,\boldsymbol{\gamma})^\alpha 
\left\{r_{\alpha,\alpha}^{(0)}(s)\lambda_0(s) - \widetilde{q}_{\alpha,\alpha}^{(0)}(s)\lambda(s,\boldsymbol{\gamma})\right\}
\right.- \Psi_{\boldsymbol{\gamma}}(s)\left\{ q_{0,\alpha}^{(0)}(s) - \widetilde{q}_{0,\alpha}^{(0)}(s)\right\}
\nonumber\\&& ~~~~~~~~~~~~~~ -\left. \lambda(s,\boldsymbol{\gamma})^\alpha \Psi_{\boldsymbol{\gamma}}(s)
\left\{r_{\alpha+1,\alpha}^{(0)}(s)\lambda_0(s) - \widetilde{q}_{\alpha+1,\alpha}^{(0)}(s)\lambda(s,\boldsymbol{\gamma})\right\}
\right]ds,
\label{EQ:MDPDE_EstEqn1M0}\\
&&\boldsymbol{u}_{0}^{(2, \alpha)}(\boldsymbol{\theta}) = 
\int_0^T \left[\lambda(s,\boldsymbol{\gamma})^\alpha 
\left\{r_{\alpha,\alpha}^{(1)}(s)\lambda_0(s) - \widetilde{q}_{\alpha,\alpha}^{(1)}(s)\lambda(s,\boldsymbol{\gamma})\right\}
\right. -\Lambda_{\boldsymbol{\gamma}}(s)\left\{ q_{0,\alpha}^{(1)}(s) - \widetilde{q}_{0,\alpha}^{(1)}(s)\right\}
\nonumber\\
&& ~~~~~~~~~~~~~~ -\left. \lambda(s,\boldsymbol{\gamma})^\alpha \Lambda_{\boldsymbol{\gamma}}(s)
\left\{r_{\alpha+1,\alpha}^{(1)}(s)\lambda_0(s) - \widetilde{q}_{\alpha+1,\alpha}^{(1)}(s)\lambda(s,\boldsymbol{\gamma})\right\}
\right]ds.
\label{EQ:MDPDE_EstEqn2M0}
\end{eqnarray}

Now, if the parametric Cox regression model (\ref{EQ:coxReg_parametric}) is indeed true, i.e, 
$\lambda_i(t) = \lambda(t, \boldsymbol{\gamma}_0)e^{\boldsymbol{\beta}_0^T\boldsymbol{z}_i}$ for all $i$
and some parameter value $\boldsymbol{\theta}_0 = (\boldsymbol{\gamma}_0^T, \boldsymbol{\beta}_0^T)^T$, 
then $\lambda_0(t) =\lambda(t, \boldsymbol{\gamma}_0)$ and 
$h_0(\boldsymbol{Z}) = e^{\boldsymbol{\beta}_0^T\boldsymbol{Z}}$ 
and hence $r_{\alpha_1,\alpha_2}^{(j)}(s) = {q}_{\alpha_1,\alpha_2}^{(j)}(s)=  \widetilde{q}_{\alpha_1,\alpha_2}^{(j)}(s)$
for each $j=0,1$ and $\alpha_1, \alpha_2\geq 0$.
Therefore, we get $\boldsymbol{u}_{0}^{(1, \alpha)}(\boldsymbol{\theta}_0) =\boldsymbol{0}$
and $\boldsymbol{u}_{0}^{(2, \alpha)}(\boldsymbol{\theta}_0) =\boldsymbol{0}$ leading to the following result.

\begin{theorem}
Suppose the observed data $(x_i, \delta_i, \boldsymbol{z}_i)$, for $i=1, \ldots, n,$ 
are IID realizations of the random variables $(X, \delta, \boldsymbol{Z})$ satisfying the assumed parametric model (\ref{EQ:coxReg_parametric}).
Then, under Assumptions (A)--(B), the MDPDE estimating equations
(\ref{EQ:MDPDE_EstEqn1})--(\ref{EQ:MDPDE_EstEqn2}) are asymptotically unbiased at the model
and the resulting MDPDE is Fisher consistent. 
\label{THM:Asymp_MDPDE1}
\end{theorem}

Whenever the model (\ref{EQ:coxReg_parametric}) does not hold exactly, 
we assume that there exists a unique solution $\boldsymbol{\theta}^g$ 
to the asymptotic estimating equations of the MDPDE
given by  $\boldsymbol{u}_{0}^{(1, \alpha)}(\boldsymbol{\theta}) =\boldsymbol{0}$
and $\boldsymbol{u}_{0}^{(2, \alpha)}(\boldsymbol{\theta}) =\boldsymbol{0}$.
We refer to this solution $\boldsymbol{\theta}^g$ as the ``best fitting parameter value";
if the model is true then $\boldsymbol{\theta}^g$ coincides with the true parameter value $\boldsymbol{\theta}_0$. 
When the model does not hold exactly, we need to make the following two assumptions
which still makes the MDPDE consistent for $\boldsymbol{\theta}^g$,
an extension of the arguments presented in \cite{Hjort:1986a, Hjort:1992}
along with the results from \citet[][p.~12-13]{Billingsley:1961}.
%

\noindent\textbf{Assumption (C):}  
There exists a neighborhood $\Theta_0$ of $ \boldsymbol{\theta}^g$
such that for every $\boldsymbol{\theta}\in \Theta_0$, 
$
- \nabla\boldsymbol{u}_{n}^{(\alpha)}({\boldsymbol{\theta}}) 
\mathop{\rightarrow}^{\mathcal{P}} J_\alpha(\boldsymbol{\theta}^g),
$ where $\nabla$ represents the gradient with respect to $\boldsymbol{\theta}$ 
and the positive definite matrix $J_\alpha(\boldsymbol{\theta}^g)$ is defined as
\begin{eqnarray}
J_\alpha(\boldsymbol{\theta})= -\nabla \begin{bmatrix}
\begin{array}{c}
\boldsymbol{u}_{0}^{(1, \alpha)}(\boldsymbol{\theta})\\
\boldsymbol{u}_{0}^{(2, \alpha)}(\boldsymbol{\theta})
\end{array}
\end{bmatrix}
= -\begin{bmatrix}
\begin{array}{cc}
\frac{\partial}{\partial\boldsymbol{\gamma}}\boldsymbol{u}_{0}^{(1, \alpha)}(\boldsymbol{\theta})
& \frac{\partial}{\partial\boldsymbol{\beta}}\boldsymbol{u}_{0}^{(1, \alpha)}(\boldsymbol{\theta})\\
\frac{\partial}{\partial\boldsymbol{\beta}}\boldsymbol{u}_{0}^{(1, \alpha)}(\boldsymbol{\theta})^T 
& \frac{\partial}{\partial\boldsymbol{\beta}}\boldsymbol{u}_{0}^{(2, \alpha)}(\boldsymbol{\theta})
\end{array}
\end{bmatrix}.
\label{EQ:J_alpha}
\end{eqnarray}

\noindent\textbf{Assumption (D):}  
There exists a neighborhood $\Theta_0$ of $ \boldsymbol{\theta}^g$ such that the quantity\\
$
K_n = \displaystyle\max_{1\leq j, k, l\leq p+q} \sup_{\boldsymbol{\theta}\in \Theta_0}\left|
\frac{1}{n} \sum_{i=1}^n  \frac{\partial^2}{\partial\theta_k\partial\theta_l} 
\boldsymbol{u}_{n, i, j}^{(\alpha)}(\boldsymbol{\theta})\right|
$
is stochastically bounded, 
where $\theta_j$ and $\boldsymbol{u}_{n, i, j}^{(\alpha)}(\boldsymbol{\theta})$ 
denote the $j$-th element of $\boldsymbol{\theta}$ and 
$\boldsymbol{u}_{n, i}^{(\alpha)}(\boldsymbol{\theta})=(\boldsymbol{u}_{n, i}^{(1, \alpha)}(\boldsymbol{\theta})^T, 
\boldsymbol{u}_{n, i}^{(2, \alpha)}(\boldsymbol{\theta}))^T$,
respectively.

Note that, one can choose the neighborhood in Assumptions (C) and (D) to be the same (otherwise choose the smaller one).
Further, these assumptions can also be verified under mild boundedness regularity conditions
in the line of \citet[][Theorem 1]{Borgan:1984}.
In the same vein, an application of martingale central limit theorem 
also yields the following condition.

\noindent\textbf{Assumption (E):}  
$
\sqrt{n}\boldsymbol{u}_{n}^{(\alpha)}({\boldsymbol{\theta}^g}) 
\displaystyle\mathop{\rightarrow}^{\mathcal{D}} N_{p+q}\left(\boldsymbol{0}_{q+p}, K_\alpha(\boldsymbol{\theta}^g)\right),
$
where 
$K_\alpha(\boldsymbol{\theta})= Var \begin{bmatrix}
\begin{array}{c}
\boldsymbol{u}_{n,i}^{(1, \alpha)}(\boldsymbol{\theta})\\
\boldsymbol{u}_{n,i}^{(2, \alpha)}(\boldsymbol{\theta})
\end{array}
\end{bmatrix}.$

Finally, Assumptions (C) and (E) can be used directly 
to derive the asymptotic distribution of the MDPDE.
Consider a Taylor series expansion of 
$\boldsymbol{u}_{n}^{(\alpha)}(\boldsymbol{\theta})= 
\left(\boldsymbol{u}_{n}^{(1, \alpha)}(\boldsymbol{\theta})^T, 
\boldsymbol{u}_{n}^{(2, \alpha)}(\boldsymbol{\theta})^T\right)^T$ 
at the MDPDE, say $\widehat{\boldsymbol{\theta}}_{n,\alpha}$, 
around the best fitting parameter value $\boldsymbol{\theta}^g$ to get
\begin{eqnarray}
\boldsymbol{0} = \boldsymbol{u}_{n}^{(\alpha)}(\widehat{\boldsymbol{\theta}}_{n,\alpha})
&=&  \boldsymbol{u}_{n}^{(\alpha)}({\boldsymbol{\theta}^g})
+ \nabla\boldsymbol{u}_{n}^{(\alpha)}(\widetilde{\boldsymbol{\theta}}) \left(\widehat{\boldsymbol{\theta}}_{n,\alpha}
-\boldsymbol{\theta}^g \right),\nonumber\\
\Rightarrow ~~~\sqrt{n} \left(\widehat{\boldsymbol{\theta}}_{n,\alpha}
-\boldsymbol{\theta}^g \right) &=&  \left[- \nabla\boldsymbol{u}_{n}^{(\alpha)}(\widetilde{\boldsymbol{\theta}})\right]^{-1}
\sqrt{n}\boldsymbol{u}_{n}^{(\alpha)}({\boldsymbol{\theta}^g}),\nonumber
\end{eqnarray}
where 
$\widetilde{\theta}$ lies between $\widehat{\boldsymbol{\theta}}_{n,\alpha}$ and $\boldsymbol{\theta}^g$.
Now using the consistency of  $\widehat{\boldsymbol{\theta}}_{n,\alpha}$ for  $\boldsymbol{\theta}^g$
and applying Assumptions (C) and (E),  we get the asymptotic distribution of the MDPDE $\widehat{\boldsymbol{\theta}}_{n,\alpha}$;
all these are summarized in the following theorem.

\begin{theorem}
Suppose the observed data $(x_i, \delta_i, \boldsymbol{z}_i)$, for $i=1, \ldots, n,$ 
are IID realizations of the random variables $(X, \delta, \boldsymbol{Z})$ having true joint distribution $H$ 
and there exists unique best fitting parameter $\boldsymbol{\theta}^g$.
Then, under Assumptions (A)--(D),  
there exists MDPDE $\widehat{\boldsymbol{\theta}}_{n,\alpha}$ 
as a solution to the estimating equations (\ref{EQ:MDPDE_EstEqn1})--(\ref{EQ:MDPDE_EstEqn2}),
which is consistent for $\boldsymbol{\theta}^g$. 
%
Also, if additionally Assumption (E) holds, 
$\sqrt{n} \left(\widehat{\boldsymbol{\theta}}_{n,\alpha} -\boldsymbol{\theta}^g \right)
\displaystyle\mathop{\rightarrow}^{\mathcal{D}}
N_{p+q}\left(\boldsymbol{0}_{q+p}, 
J_\alpha(\boldsymbol{\theta}^g)^{-1} K_\alpha(\boldsymbol{\theta}^g)J_\alpha(\boldsymbol{\theta}^g)^{-1}\right)$.
\label{THM:Asymp_MDPDE}
\end{theorem}

The next theorem presents a consistent estimate of the asymptotic variance matrix of the MDPDE
which can be used for estimating their standard errors and 
for developing robust significance testing procedures based on these MDPDEs.
The proof follows from martingale inequalities and uniform convergence in probability arguments; 
see \cite{Hjort:1991a, Hjort:1992} for a similar argument in case of the MLE.

\begin{theorem}
\label{THM:Asymp_MDPDE_VarEst}
Under the assumptions of Theorem \ref{THM:Asymp_MDPDE}, 
a consistent estimate of the asymptotic  variance of the MDPDE $\widehat{\boldsymbol{\theta}}_{n,\alpha}
=\left(\widehat{\boldsymbol{\gamma}}_{n,\alpha}^T, \widehat{\boldsymbol{\beta}}_{n,\alpha}^T\right)^T$
is given by $\frac{1}{n}\widehat{J}_{n,\alpha}^{-1} \widehat{K}_{n,\alpha}\widehat{J}_{n,\alpha}^{-1}$,
where $\widehat{J}_{n,\alpha}$ and $\widehat{K}_{n,\alpha}$ are consistent estimates of the matrices 
$J_\alpha(\boldsymbol{\theta}^g)$ and $K_\alpha(\boldsymbol{\theta}^g)$, respectively, given by 

\begin{eqnarray}
\widehat{J}_{n,\alpha}= - \nabla\boldsymbol{u}_{n}^{(\alpha)}(\widehat{\boldsymbol{\theta}}_{n,\alpha}),
~~~\mbox{ and }~~~~ 
\widehat{K}_{n,\alpha} = \frac{1}{n}\sum_{i=1}^n \begin{bmatrix}
\begin{array}{cc}
\widehat{L}_{1,\alpha,i}\widehat{L}_{1, \alpha,i}^T
& \widehat{L}_{1,\alpha,i}\widehat{L}_{2, \alpha,i}^T\\
\widehat{L}_{2,\alpha,i}\widehat{L}_{1, \alpha,i}^T 
& \widehat{L}_{2,\alpha,i}\widehat{L}_{2, \alpha,i}^T
\end{array}
\end{bmatrix},
\end{eqnarray}
with $\widehat{S}_i(x_i) = \exp\left[
-\Lambda(x_i,\widehat{\boldsymbol{\gamma}}_{n,\alpha})e^{\widehat{\boldsymbol{\beta}}_{n,\alpha}^T\boldsymbol{z}_i}\right]$
and 
\begin{eqnarray}
\widehat{L}_{1,\alpha,i} &=& 
\left[\left\{ \psi_{\widehat{\boldsymbol{\gamma}}_{n,\alpha}}(x_i)\lambda(x_i,\widehat{\boldsymbol{\gamma}}_{n,\alpha})^\alpha 
e^{\alpha\widehat{\boldsymbol{\beta}}_{n,\alpha}^T\boldsymbol{z}_i}\widehat{S}_i(x_i)^\alpha 
\right.\right.
\nonumber\\
&& ~~~~~~ -\left.\left(\lambda(x_i,\widehat{\boldsymbol{\gamma}}_{n,\alpha})^\alpha 
e^{\alpha\widehat{\boldsymbol{\beta}}_{n,\alpha}^T\boldsymbol{z}_i}-1\right)
\Psi_{\widehat{\boldsymbol{\gamma}}_{n,\alpha}}(x_i)e^{\widehat{\boldsymbol{\beta}}_{n,\alpha}^T\boldsymbol{z}_i}
\widehat{S}_i(x_i)^\alpha\right\}\delta_i
\nonumber\\
&& ~~~~~~~~~ -\left. 
\Psi_{\widehat{\boldsymbol{\gamma}}_{n,\alpha}}(x_i)e^{\widehat{\boldsymbol{\beta}}_{n,\alpha}^T\boldsymbol{z}_i}\widehat{S}_i(x_i)^\alpha 
-\xi_{i}^{(1,\alpha)}(\widehat{\boldsymbol{\theta}}_{n,\alpha})\right],
\label{EQ:L_alphai1}\\
\widehat{L}_{2,\alpha,i} &=& 
\left[\left\{ \lambda(x_i,\widehat{\boldsymbol{\gamma}}_{n,\alpha})^\alpha 
e^{\alpha\widehat{\boldsymbol{\beta}}_{n,\alpha}^T\boldsymbol{z}_i}\widehat{S}_i(x_i)^\alpha
\right.\right.
\nonumber\\
&& ~~~~~~ -\left.\left(\lambda(x_i,\widehat{\boldsymbol{\gamma}}_{n,\alpha})^\alpha 
e^{\alpha\widehat{\boldsymbol{\beta}}_{n,\alpha}^T\boldsymbol{z}_i}-1\right)\Lambda_{\widehat{\boldsymbol{\gamma}}_{n,\alpha}}(x_i)
e^{\widehat{\boldsymbol{\beta}}_{n,\alpha}^T\boldsymbol{z}_i}\widehat{S}_i(x_i)^\alpha\right\}
\boldsymbol{z}_i\delta_i
\nonumber\\
&& ~~~~~~~~~~~ 
-\left. \boldsymbol{z}_i\Lambda_{\widehat{\boldsymbol{\gamma}}_{n,\alpha}}(x_i)e^{\boldsymbol{\beta}^T\boldsymbol{z}_i}\widehat{S}_i(x_i)^\alpha 
-\xi_{i}^{(2,\alpha)}(\widehat{\boldsymbol{\theta}}_{n,\alpha})\right].
\label{EQ:L_alphai2}
\end{eqnarray}
\end{theorem}

\begin{remark}
At $\alpha=0$, the MDPDE coincides with the MLE and hence our Theorems \ref{THM:Asymp_MDPDE} and \ref{THM:Asymp_MDPDE_VarEst}
yield the asymptotic properties of the MLE as a special case, which coincide with the earlier results in Theorem 6.1 of \cite{Hjort:1992}.
In particular, our matrices $J_\alpha$ and $K_\alpha$ at $\alpha=0$ coincide with the matrices $J$ and $K$ of \citet[p.~377]{Hjort:1992}. 
\end{remark}

\section{Robustness: Influence Function Analysis}\label{SEC:robust}

We now study the robustness of the proposed MDPDE theoretically 
through the classical influence function analysis \citep{Hampel/etc:1986}.
%
In the context of censored data, the concept of  influence function (IF) has been extended by 
\cite{Reid:1981}, \cite{Reid/Crepeau:1985} and \cite{Hjort:1992} among others. 
In particular, \cite{Hjort:1992} derived the IF of the MLE of the parameters in the Cox regression model (\ref{EQ:coxReg_parametric})
and argued that it is unbounded indicating the non-robust nature of the MLE. 
Here we derive the IF of the proposed MDPDE at $\alpha>0$.

Note that the MDPDE functional at any $\alpha \geq 0$, say $\boldsymbol{T}_\alpha(H)=\boldsymbol{\theta}^g$,
at the true distribution $H$ of the triplet $(X, \delta, \boldsymbol{Z})$ can be defined as the solution to the 
asymptotic estimating equations $\boldsymbol{u}_{0}^{(1, \alpha)}(\boldsymbol{\theta}) =\boldsymbol{0}$
and $\boldsymbol{u}_{0}^{(2, \alpha)}(\boldsymbol{\theta}) =\boldsymbol{0}$.
Now, let us consider a point mass contamination at the point $(x_t, \delta_t, \boldsymbol{z}_t)$
and the contaminated density $H_\epsilon = (1-\epsilon)H + \epsilon\wedge_{(x_t, \delta_t, \boldsymbol{z}_t)}$
where $\wedge$ denotes a degenerate distribution. 
Then, the IF of the proposed MDPDE is defined as 
\begin{equation}
IF((x_t, \delta_t, \boldsymbol{z}_t); \boldsymbol{T}_\alpha, H) = \lim\limits_{\epsilon\downarrow 0} 
\frac{\boldsymbol{T}_\alpha(H_\epsilon) - \boldsymbol{T}_\alpha(H)}{\epsilon} =
\left. \frac{\partial}{\partial\epsilon}\boldsymbol{T}_\alpha(H_\epsilon)\right|_{\epsilon=0}.\nonumber
\end{equation}
In order to derive this IF, we substitute $\boldsymbol{T}_\alpha(H_\epsilon)$ and $H_\epsilon$ in place of $\boldsymbol{T}_\alpha(H)$
and $H$, respectively, in the asymptotic estimating equations and then differentiate with respect to $\epsilon$ at $\epsilon=0$. 
Collecting terms, after some lengthy but routine algebra, the IF of the MDPDE becomes 
\begin{eqnarray}
&&IF((x_t, \delta_t, \boldsymbol{z}_t); \boldsymbol{T}_\alpha, H) = J_\alpha(\boldsymbol{\theta}^g)^{-1}
\begin{bmatrix}
\begin{array}{c}
\boldsymbol{i}_{1,\alpha}((x_t, \delta_t, \boldsymbol{z}_t); \boldsymbol{\theta}^g)\\
\boldsymbol{i}_{2,\alpha}((x_t, \delta_t, \boldsymbol{z}_t); \boldsymbol{\theta}^g)
\end{array}
\end{bmatrix},
\label{EQ:IF_MDPDE}\\
\mbox{where}
&&\boldsymbol{i}_{1,\alpha}((x_t, \delta_t, \boldsymbol{z}_t); \boldsymbol{\theta}) = 
-\Psi_{{\boldsymbol{\gamma}}}(x_t)e^{{\boldsymbol{\beta}}^T\boldsymbol{z}_t}{S}_t(x_t)^\alpha 
-\xi_{t}^{(1,\alpha)}({\boldsymbol{\theta}})
\nonumber\\
&& ~~~~+\left\{ \psi_{{\boldsymbol{\gamma}}}(x_t)\lambda(x_t,{\boldsymbol{\gamma}})^\alpha 
e^{\alpha{\boldsymbol{\beta}}^T\boldsymbol{z}_t}{S}_t(x_t)^\alpha 
-\left(\lambda(x_t,{\boldsymbol{\gamma}})^\alpha 
e^{\alpha{\boldsymbol{\beta}}^T\boldsymbol{z}_t}-1\right)
\Psi_{{\boldsymbol{\gamma}}}(x_t)e^{{\boldsymbol{\beta}}^T\boldsymbol{z}_t}
{S}_t(x_t)^\alpha\right\}\delta_t,
\nonumber\label{EQ:i_alphai1}\\
&&\boldsymbol{i}_{1,\alpha}((x_t, \delta_t, \boldsymbol{z}_t); \boldsymbol{\theta}) = 
- \boldsymbol{z}_t\Lambda_{{\boldsymbol{\gamma}}}(x_t)e^{\boldsymbol{\beta}^T\boldsymbol{z}_t}{S}_t(x_t)^\alpha 
-\xi_{t}^{(2,\alpha)}({\boldsymbol{\theta}})
\nonumber\\
&& ~~~~~~ + \left\{ \lambda(x_t,{\boldsymbol{\gamma}})^\alpha 
e^{\alpha{\boldsymbol{\beta}}^T\boldsymbol{z}_t}{S}_t(x_t)^\alpha
 -\left(\lambda(x_t,{\boldsymbol{\gamma}})^\alpha 
e^{\alpha{\boldsymbol{\beta}}^T\boldsymbol{z}_t}-1\right)\Lambda_{{\boldsymbol{\gamma}}}(x_t)
e^{{\boldsymbol{\beta}}^T\boldsymbol{z}_t}{S}_t(x_t)^\alpha\right\}
\boldsymbol{z}_t\delta_t.
\nonumber
\label{EQ:i_alphai2}
\end{eqnarray}
Note that, due to the presence of the terms $\lambda(x_t,{\boldsymbol{\gamma}})^\alpha$,  
$e^{\alpha{\boldsymbol{\beta}}^T\boldsymbol{z}_t}$ and $S_t(x_t)^\alpha$, the above IF of the MDPDE 
remains bounded over contamination points at any $\alpha>0$. This implies the claimed robustness properties of the MDPDEs with $\alpha>0$.

Further, a diagnostic measure for the $i$-th observation can be obtained from this IF as 
$
\widehat{I}_i = IF((x_i, \delta_i, \boldsymbol{z}_i); \boldsymbol{T}_\alpha, \widehat{H}_n) 
= \widehat{J}_{n, \alpha}^{-1}
\begin{bmatrix}
\begin{array}{c}
\widehat{L}_{1,\alpha,i}\\
\widehat{L}_{2,\alpha, i}
\end{array}
\end{bmatrix}
= \widehat{J}_{n, \alpha}^{-1}
\begin{bmatrix}
\begin{array}{c}
\boldsymbol{i}_{1,\alpha}((x_i, \delta_i, \boldsymbol{z}_i); \widehat{\boldsymbol{\theta}}_{n,\alpha})\\
\boldsymbol{i}_{2,\alpha}((x_i, \delta_i, \boldsymbol{z}_i); \widehat{\boldsymbol{\theta}}_{n,\alpha})
\end{array}
\end{bmatrix},
$ \\
where $\widehat{H}_n$ is the empirical distribution of the observed data $(x_i, \delta_i, \boldsymbol{z}_i), i=1, \ldots, n$.
Larger values of $\widehat{I}_i$ indicate greater influence of the $i$-th observations in computing the MDPDEs
and hence it tends to be an outlying observation.

\section{Simulation Studies}\label{SEC:simulation}

Here we present a few interesting findings from some simulation studies  
in order to examine the finite sample properties of the proposed MDPDE. 
%
Consider the exponential baseline hazard in the parametric Cox regression model (\ref{EQ:coxReg_parametric})
given by $\lambda(t, \gamma) = \gamma \in [0, \infty)$. 
Here our parameter of interest 
is the $p+1$ dimensional vector $(\gamma, \boldsymbol{\beta}^T)^T$. 
We simulate samples of size $n$ from this model 
with the covariates being generated from standard normal distributions;
uniform random censoring with censoring proportions 5\% or 10\% are applied to these data. 
We report the results for the specific case with $p=2$ where the true parameter values 
are taken as $\gamma_0=1$ and $\boldsymbol{\beta}_0 = (2, -2)^T$.
To study the effect of contaminations, $100\epsilon\%$ of each sample is contaminated 
by replacing the covariate values by IID observations from $N(1,4)$ distribution.
We then compute the MDPDEs of the three parameters $(\gamma, \beta_1, \beta_2)^T$ 
for different $\alpha\geq 0$ ($\alpha=0$ generates the MLE) based on each simulated sample.
The whole process is repeated over 1000 samples to compute the empirical biases and MSEs of the MDPDEs in all cases, 
which 
are reported in Tables \ref{TAB:MDPDE_Bias_n50}--\ref{TAB:MDPDE_MSE_n50} for $n=50$ and $100$.
For comparison, we also report the empirical bias and MSE
of the partial likelihood estimator (PLE) of \cite{Cox:1975}
and the popular robust estimator of \cite{Bednarski:1993} (BRE) for $\boldsymbol{\beta}=(\beta_1, \beta_2)^T$ 
with the semi-parametric Cox model (\ref{EQ:coxReg}) for the same sets of simulated data; 
these PLE and BRE are computed using the R package `coxrobust' \citep{Bednarski/Borowicz:2006}.

It is clear from the tables that both the parametric MLE (MDPDE at $\alpha=0$)
and the semi-parametric MLE (PLE) are highly non-robust against any amount of contamination in data.
Under pure data, the parametric MLE has the least bias and MSE in all cases
and the empirical bias and MSE of the MDPDEs increase as $\alpha$ increases.
However, the loss in efficiency under pure data
is not very significant for the MDPDEs with small $\alpha>0$; 
for most $\alpha$ they are, in fact, still smaller than 
the semi-parametric PLE and its existing robust version BRE.
In other words, if the parametric model is correctly specified without contamination, 
the MDPDEs with small $\alpha>0$ yield the least (insignificant) loss in efficiency 
among other existing robust estimators when compared to the fully efficient (but non-robust) MLE. 

On the other hand, when there is some amount of contamination in the data, 
the bias and the MSE of the parametric MLE increase significantly, 
but those of the MDPDEs remain more stable at $\alpha>0$. 
In particular, as $\alpha$ increases, both bias and MSE of the MDPDEs initially have a substantial 
decrease under data contamination but then increase again at larger $\alpha$ values 
(due to their low efficiency in pure data).
From the tables, one can see that the MDPDEs with $\alpha\in (0.2, 0.4)$ 
give the least bias and MSE under contamination and these are lower than the same 
for the existing robust estimator (BRE) with semi-parametric modeling.  
Considering their pure data performance, the MDPDEs with 
 $\alpha\in (0.2, 0.4)$   yield the best trade-off between the efficiency and robustness
for estimating the parameters under the Cox regression model. 
This parametric formulation also makes it possible to estimate 
the baseline hazard function through robust and efficient estimation of the underlying parameter $\boldsymbol{\gamma}$.
These clearly illustrate the advantages of the proposed MDPDE 
under the parametric Cox regression model for more precise and robust estimation
in the cases where data may be prone to outlying observations.

\section{Real Data Applications}\label{SEC:real_data}

We will now apply the proposed parametric Cox regression model to analyze 
two interesting survival data examples in the context of medical science.
Both examples are seen to produce incorrect results,
when analyzed using the usual semi-parametric Cox models using the R package `coxrobust' \citep{Bednarski/Borowicz:2006}, 
due to the presence of a few outliers.
For brevity, we will not present these detailed semi-parametric analysis
and illustrate the advantages of our proposal by fitting a suitable parametric Cox regression model
with only the significant covariates obtained from the semi-parametric full model analysis.
The parametric baseline hazards are chosen individually for each dataset from the plot of 
the cumulative baseline hazard estimated from semi-parametric analysis
and the corresponding Cox-Snell residual plot is used to identify the outliers (see Figure \ref{FIG:Data}).
The MLE and the MDPDE of all the parameters are compared for data with and without these outliers,
which clearly exhibits significantly better stability of the proposed MDPDEs compared to the MLE.

\subsection{Myeloma Data}

The first example is the survival data of 65 multiple myeloma patients obtained from \cite{Heritier/etc:2009},
where the associated significant covariates  are 
the logarithms of blood urea nitrogen (BUN), serum calcium at diagnosis (CALC) and hemoglobin (HGB) of the patients.
Further, the survival times of only 48 patients are observed and 
that for the other 17 patients are right censored;
so the censoring proportion in the data is quite high, approximately  26\%.

\begin{figure}[h!]
	\centering
	\subfloat[Myeloma Data]{
		\includegraphics[width=0.5\textwidth]{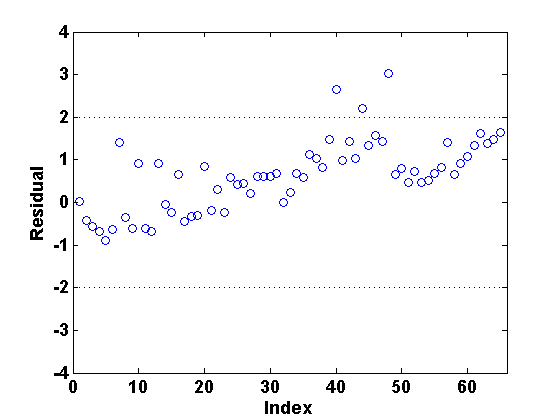}
		\includegraphics[width=0.5\textwidth]{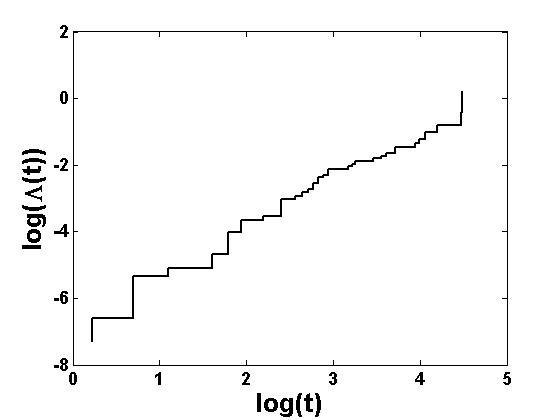}
		\label{FIG:Data_Myeloma}}
\\ 
	\subfloat[Breast and Ovarian Cancer Data]{
		\includegraphics[width=0.5\textwidth]{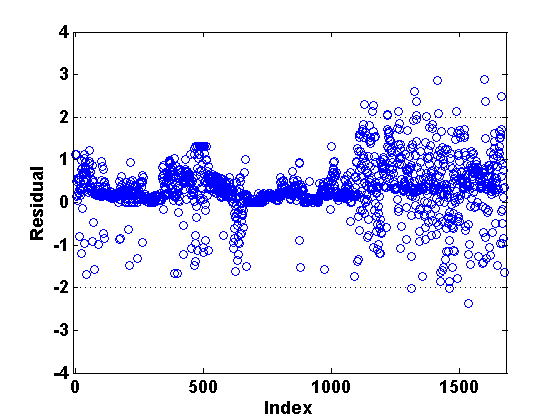}
\includegraphics[width=0.5\textwidth]{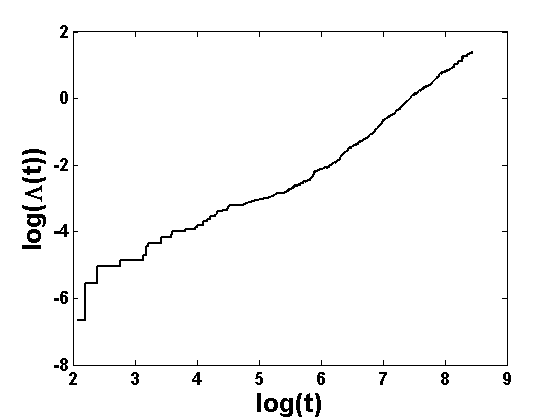}
		\label{FIG:Data_BRCAOV}}
	\caption{Residual plots (left) and the empirical estimate of the cumulative hazard $\Lambda(t)$ (right) 
		for the two real datasets}
	\label{FIG:Data}
\end{figure}

As mentioned above, the usual semi-parametric Cox model is initially fitted and 
the resulting Cox-Snell residuals based on the deviance method, presented in Figure \ref{FIG:Data_Myeloma}, 
reveal three  outlying points in the dataset having deviance values  outside the 95\% tolerance range $[-2, 2]$.
Further, the corresponding estimate of the cumulative baseline hazard function is plotted in the log-scale 
in Figure \ref{FIG:Data_Myeloma}. It is clearly evident from the figure that 
the  cumulative baseline hazard closely resembles a $y=x$ straight line in the log scale; 
this particular form corresponds to the exponential hazard function 
$\lambda(t, \boldsymbol{\gamma}) = \gamma$, for a constant $\gamma>0$.
So, we fit the parametric Cox regression model (\ref{EQ:coxReg_parametric})
with the above exponential baseline hazard and the previously mentioned three covariates (BUN, CALC and HGB).
Under this fully parametric set-up, we compute the estimates of the regression coefficients
and the parameters $\boldsymbol{\gamma}$ in the hazard function using the MLE and 
the proposed MDPDEs with different $\alpha>0$; these are reported in Table \ref{TAB:Data_Myeloma}. 
However, due to the presence of the outliers, the MLE gets affected significantly.  
To demonstrate this, we re-compute the MLE and the MDPDEs  after removing the three outlying observations from the dataset
which are also reported in Table \ref{TAB:Data_Myeloma}.
One can clearly see that the MDPDEs remain much more stable in the presence of outliers 
compared to the MLE under the fully parametric Cox regression model.
In particular the MDPDEs at $\alpha=0.3, 0.4$ and $0.5$ show excellent stability. 
Additionally, this parametric version gives us the flexibility to estimate the baseline hazard
in a more rigorous parametric form.

\subsection{Breast and Ovarian Cancer Data}

Our second example is from a large Breast and Ovarian cancer trial 
with 1619 patients' suffering either from breast or ovarian cancer;
the corresponding number of patients in the two types of cancer are 1044 and 575, respectively.
These data are obtained from a recent R-package \textit{survminer}, 
after filtering out the erroneous observations of zero and negative lifetimes. 
In total, the lifetime of 401 patients are not observed exactly, yielding a censoring proportion of approximately 24\%.
We want to get an idea about the difference in patient's lifetimes between the two types of cancer;
this can be achieved through a Cox type modeling with only one indicator covariate (say ``Type")
which we take to be one for breast cancer.

Again, we first apply the standard semi-parametric Cox model;
the resulting Cox-Snell residuals based on the deviance method
and the estimate of the cumulative baseline hazard function are presented Figure \ref{FIG:Data_BRCAOV}.
Note that, 19 observations have residual values outside the 95\% tolerance range $[-2, 2]$
and are identified  as outliers. Further, the  cumulative baseline hazard can be 
well approximated by a straight line in the log scale, which leads to the Weibull hazard function given by
$
\lambda(t, \boldsymbol{\gamma}) = \gamma_1\gamma_2 t^{\gamma_2-1}$, 
for $\boldsymbol{\gamma}=(\gamma_1, \gamma_2)^T\in [0, \infty)^2$.
Therefore, we apply the proposed MDPDE, along with MLE, to fit 
the parametric Cox regression model (\ref{EQ:coxReg_parametric}) 
with the above Weibull baseline hazard function and only one covariate ``Type" (of cancer).
The parameter estimates obtained under the full data and the outlier deleted data are reported in Table \ref{TAB:Data_BRCAOV}. 
Once again, it is clearly observed that the MDPDEs with larger $\alpha>0$ 
are far more stable than the MLE. 
However it may be noted that these MDPDEs are ot necessarily close to the outlier deleted MLE.
This is an outcome of the fact that the outlier detection based on the Cox-Snell residuals is far too liberal a process in this case
and identifies too few outliers relative to the robust MDPDE procedure.
The situation changes if the trimming proportion is increased. 
For example, trimming 15\% of the extreme residuals (in absolute values)
pushes the outlier deleted MLE to the neighborhood of the full data MDPDE at $\alpha=0.3$.
On the whole it is obvious that the MDPDE gives good and stable inference, even under the full data, 
in this example.

\section{Concluding Remarks}

This paper presents a fully parametric alternative to the semi-parametric Cox model
for more precise and efficient inference under randomly censored responses. 
Due to the non-robust nature of the existing maximum likelihood approach under data contamination,
we here develop a robust generalization, 
namely the robust technique using the minimum density power divergence estimator (MDPDE), 
which provides better trade-off between efficiency and robustness under the fully parametric Cox  regression model.
In particular, we have illustrated that the MDPDEs with tuning parameter $\alpha \in (0.2, 0.4)$
generate highly robust estimators in the presence of contamination while
there is no significant loss in their efficiency under pure data. 
Therefore, these MDPDEs can be used in practice to get better and stable inference in analyzing 
different practical datasets which are prone to the presence of outlying observations.
We have also derived in brief the asymptotic properties of the proposed MDPDE under 
the fully parametric Cox regression model to show its consistency and asymptotic normality. 
We also provide a consistent estimate of the asymptotic variance matrix  of the MDPDEs
which can be used to estimate their standard errors in any practical applications.

There could be several extensions of this work with interesting real life applications. 
The asymptotic variances of the MDPDEs and their estimates can later be used to develop 
robust hypothesis testing or model selection procedures under the fully parametric Cox regression model.
The concept of efficient parametric  formulations and robust estimation using the MDPDEs
can also be extended to many different applications involving 
the semi-parametric or non-parametric counting process models with or without censoring.
Finally, although some empirical suggestions are made for $\alpha$ providing best trade-offs,  
a data driven choice of this MDPDE tuning parameter  $\alpha$, 
along the lines of \cite{Warwick/Jones:2005} and \cite{Ghosh/Basu:2015}, would be really helpful 
for practitioners from applied sciences to use the proposed procedure. 
Further, in our second example, the limitation of the Cox-Snell residual approach in identifying 
all the outliers  in a contaminated dataset is clearly observed
and hence a new robust version of the Cox-Snell residual, possibly based on the proposed MDPDEs,
 will be more helpful and effective for outlier detection.
We hope to pursue some of these extensions in our future research.

\bigskip\noindent\textbf{Acknowledgment:}
The research of the first author (AG) is partially supported by the INSPIRE Faculty Research Grant
from  Department of Science and Technology, Government of India.


\begin{table}[h]
	\centering
	\caption{Empirical biases of the estimates of $\boldsymbol{\beta}=(\beta_1, \beta_2)^T$
		and $\gamma$ for $n=50$ and 100 and different contamination proportion $\epsilon$ }
	\resizebox{.85\textwidth}{!}{
		\begin{tabular}{lll|rrrrrrr|rr}\hline
			Cens.	&	&   & \multicolumn{7}{c|}{Parametric MDPDE with $\alpha$} & \multicolumn{2}{c}{Semi-parametric}\\		
			Prop.	&	$\epsilon$	&		&	0 (MLE)	&	0.05	&	0.1	&	0.2	&	0.3	&	0.4	&	0.5	&	PLE	&	BRE	\\\hline
		\multicolumn{12}{c}{$n=50$}\\\hline
			5\% 	&	0	&	$\beta_1$	&	$-$0.011	&	0.001	&	0.011	&	0.034	&	0.064	&	0.115	&	0.200	&	0.060	&	0.212	\\
			&		&	$\beta_2$	&	0.000	&	$-$0.011	&	$-$0.021	&	$-$0.045	&	$-$0.078	&	$-$0.119	&	$-$0.190	&	$-$0.073	&	$-$0.220	\\
			&		&	$\gamma$	&	$-$0.014	&	$-$0.024	&	$-$0.035	&	$-$0.059	&	$-$0.089	&	$-$0.121	&	$-$0.150	&	--	&	--	\\\\
			&	0.05	&	$\beta_1$	&	$-$0.704	&	$-$0.401	&	$-$0.166	&	$-$0.086	&	$-$0.108	&	$-$0.144	&	$-$0.137	&	$-$0.928	&	$-$0.338	\\
			&		&	$\beta_2$	&	0.254	&	0.104	&	0.019	&	$-$0.014	&	$-$0.035	&	$-$0.067	&	$-$0.186	&	0.706	&	0.251	\\
			&		&	$\gamma$	&	$-$0.397	&	$-$0.211	&	$-$0.091	&	$-$0.061	&	$-$0.063	&	$-$0.045	&	$-$0.008	&	--	&	--	\\\\
			&	0.1	&	$\beta_1$	&	$-$0.888	&	$-$0.558	&	$-$0.240	&	$-$0.150	&	$-$0.190	&	$-$0.259	&	$-$0.296	&	$-$1.193	&	$-$0.692	\\
			&		&	$\beta_2$	&	0.551	&	0.253	&	0.098	&	0.045	&	0.025	&	0.018	&	$-$0.064	&	1.050	&	0.682	\\
			&		&	$\gamma$	&	$-$0.515	&	$-$0.289	&	$-$0.112	&	$-$0.054	&	$-$0.030	&	0.025	&	0.101	&	--	&	--	\\
			\hline\\
			10\% 	&	0	&	$\beta_1$	&	$-$0.001	&	0.016	&	0.034	&	0.071	&	0.109	&	0.172	&	0.250	&	0.078	&	0.242	\\
			&		&	$\beta_2$	&	0.002	&	$-$0.014	&	$-$0.032	&	$-$0.068	&	$-$0.104	&	$-$0.167	&	$-$0.232	&	$-$0.083	&	$-$0.242	\\
			&		&	$\gamma$	&	$-$0.071	&	$-$0.087	&	$-$0.105	&	$-$0.145	&	$-$0.189	&	$-$0.234	&	$-$0.260	&	--	&	--	\\\\
			&	0.05	&	$\beta_1$	&	$-$0.745	&	$-$0.439	&	$-$0.178	&	$-$0.081	&	$-$0.115	&	$-$0.158	&	$-$0.152	&	$-$0.967	&	$-$0.358	\\
			&		&	$\beta_2$	&	0.274	&	0.105	&	0.021	&	$-$0.021	&	$-$0.038	&	$-$0.083	&	$-$0.182	&	0.733	&	0.261	\\
			&		&	$\gamma$	&	$-$0.445	&	$-$0.276	&	$-$0.162	&	$-$0.140	&	$-$0.152	&	$-$0.154	&	$-$0.126	&	--	&	--	\\\\
			&	0.1	&	$\beta_1$	&	$-$0.934	&	$-$0.561	&	$-$0.274	&	$-$0.151	&	$-$0.193	&	$-$0.244	&	$-$0.259	&	$-$1.219	&	$-$0.686	\\
			&		&	$\beta_2$	&	0.586	&	0.275	&	0.105	&	0.024	&	0.010	&	$-$0.024	&	$-$0.078	&	1.063	&	0.664	\\
			&		&	$\gamma$	&	$-$0.534	&	$-$0.331	&	$-$0.185	&	$-$0.142	&	$-$0.141	&	$-$0.110	&	$-$0.055	&	--	&	--	\\	
			\hline
		\end{tabular}
	}
	\resizebox{.85\textwidth}{!}{
	\begin{tabular}{lll|rrrrrrr|rr}\hline
		\multicolumn{12}{c}{$n=100$}\\\hline
		5\%&	0	&	$\beta_1$	&	$-$0.006	&	0.000	&	0.007	&	0.021	&	0.041	&	0.067	&	0.128	&	0.029	&	0.107	\\
		&		&	$\beta_2$	&	$-$0.001	&	$-$0.009	&	$-$0.017	&	$-$0.034	&	$-$0.055	&	$-$0.087	&	$-$0.169	&	$-$0.038	&	$-$0.117	\\
		&		&	$\gamma$	&	$-$0.030	&	$-$0.038	&	$-$0.047	&	$-$0.066	&	$-$0.089	&	$-$0.117	&	$-$0.145	&	--	&	--	\\\\
		&	0.05	&	$\beta_1$	&	$-$0.809	&	$-$0.208	&	$-$0.058	&	$-$0.031	&	$-$0.066	&	$-$0.125	&	$-$0.164	&	$-$1.055	&	$-$0.338	\\
		&		&	$\beta_2$	&	0.431	&	0.085	&	0.021	&	0.007	&	0.009	&	0.003	&	$-$0.026	&	0.889	&	0.339	\\
		&		&	$\gamma$	&	$-$0.459	&	$-$0.138	&	$-$0.064	&	$-$0.058	&	$-$0.054	&	$-$0.024	&	0.035	&	--	&	--	\\\\
		&	0.1	&	$\beta_1$	&	$-$1.182	&	$-$0.439	&	$-$0.167	&	$-$0.136	&	$-$0.219	&	$-$0.342	&	$-$0.454	&	$-$1.427	&	$-$0.726	\\
		&		&	$\beta_2$	&	0.816	&	0.209	&	0.072	&	0.045	&	0.062	&	0.095	&	0.100	&	1.264	&	0.710	\\
		&		&	$\gamma$	&	$-$0.618	&	$-$0.236	&	$-$0.082	&	$-$0.048	&	$-$0.012	&	0.070	&	0.205	&	--	&	--	\\
		\hline\\
		10\%	&	0	&	$\beta_1$	&	$-$0.007	&	0.008	&	0.023	&	0.053	&	0.080	&	0.112	&	0.160	&	0.028	&	0.110	\\
		&		&	$\beta_2$	&	$-$0.003	&	$-$0.017	&	$-$0.032	&	$-$0.061	&	$-$0.090	&	$-$0.136	&	$-$0.216	&	$-$0.037	&	$-$0.116	\\
		&		&	$\gamma$	&	$-$0.079	&	$-$0.092	&	$-$0.107	&	$-$0.141	&	$-$0.179	&	$-$0.219	&	$-$0.248	&	--	&	--	\\\\
		&	0.05	&	$\beta_1$	&	$-$0.801	&	$-$0.215	&	$-$0.069	&	$-$0.026	&	$-$0.059	&	$-$0.112	&	$-$0.154	&	$-$1.057	&	$-$0.331	\\
		&		&	$\beta_2$	&	0.452	&	0.070	&	0.008	&	$-$0.036	&	$-$0.053	&	$-$0.078	&	$-$0.095	&	0.894	&	0.321	\\
		&		&	$\gamma$	&	$-$0.492	&	$-$0.201	&	$-$0.139	&	$-$0.144	&	$-$0.161	&	$-$0.152	&	$-$0.113	&	--	&	--	\\\\
		&	0.1	&	$\beta_1$	&	$-$1.169	&	$-$0.466	&	$-$0.169	&	$-$0.109	&	$-$0.193	&	$-$0.321	&	$-$0.430	&	$-$1.412	&	$-$0.706	\\
		&		&	$\beta_2$	&	0.823	&	0.230	&	0.070	&	0.020	&	0.034	&	0.067	&	0.041	&	1.257	&	0.698	\\
		&		&	$\gamma$	&	$-$0.642	&	$-$0.311	&	$-$0.165	&	$-$0.137	&	$-$0.124	&	$-$0.057	&	0.042	&	--	&	--	\\
		\hline
	\end{tabular}
}	\label{TAB:MDPDE_Bias_n50}
\end{table}

\begin{table}[h]
	\centering
	\caption{Empirical MSEs of the estimates of $\boldsymbol{\beta}=(\beta_1, \beta_2)^T$
		and $\gamma$ for $n=50$ and 100 and different contamination proportion $\epsilon$ }
	\resizebox{.85\textwidth}{!}{
		\begin{tabular}{lll|rrrrrrr|rr}\hline
			Cens.	&	&   & \multicolumn{7}{c|}{Parametric MDPDE with $\alpha$} & \multicolumn{2}{c}{Semi-parametric}\\		
			Prop.	&	$\epsilon$	&		&	0 (MLE)	&	0.05	&	0.1	&	0.2	&	0.3	&	0.4	&	0.5	&	PLE	&	BRE	\\\hline
		\multicolumn{12}{c}{$n=50$}\\\hline
			5\% 	&	0	&	$\beta_1$	&	0.025	&	0.026	&	0.030	&	0.045	&	0.095	&	0.291	&	0.822	&	0.103	&	0.215	\\
			&		&	$\beta_2$	&	0.026	&	0.028	&	0.032	&	0.050	&	0.119	&	0.267	&	0.762	&	0.116	&	0.227	\\
			&		&	$\gamma$	&	0.022	&	0.023	&	0.024	&	0.030	&	0.048	&	0.087	&	0.160	&	--	&	--	\\\\
			&	0.05	&	$\beta_1$	&	0.862	&	0.433	&	0.165	&	0.120	&	0.207	&	0.369	&	0.855	&	1.206	&	0.326	\\
			&		&	$\beta_2$	&	0.371	&	0.180	&	0.077	&	0.077	&	0.154	&	0.334	&	1.344	&	0.762	&	0.266	\\
			&		&	$\gamma$	&	0.289	&	0.126	&	0.051	&	0.041	&	0.071	&	0.173	&	0.380	&	--	&	--	\\\\
			&	0.1	&	$\beta_1$	&	1.182	&	0.652	&	0.251	&	0.166	&	0.246	&	0.485	&	0.923	&	1.672	&	0.676	\\
			&		&	$\beta_2$	&	0.701	&	0.308	&	0.132	&	0.102	&	0.205	&	0.482	&	1.455	&	1.307	&	0.647	\\
			&		&	$\gamma$	&	0.415	&	0.183	&	0.071	&	0.048	&	0.068	&	0.186	&	0.445	&	--	&	--	\\
			\hline\\
			10\% 	&	0	&	$\beta_1$	&	0.025	&	0.028	&	0.032	&	0.055	&	0.108	&	0.341	&	0.961	&	0.129	&	0.261	\\
			&		&	$\beta_2$	&	0.026	&	0.028	&	0.033	&	0.054	&	0.098	&	0.362	&	0.751	&	0.125	&	0.248	\\
			&		&	$\gamma$	&	0.029	&	0.031	&	0.035	&	0.049	&	0.074	&	0.124	&	0.189	&	--	&	--	\\\\
			&	0.05	&	$\beta_1$	&	0.946	&	0.501	&	0.199	&	0.122	&	0.202	&	0.394	&	0.903	&	1.283	&	0.353	\\
			&		&	$\beta_2$	&	0.409	&	0.172	&	0.078	&	0.074	&	0.127	&	0.363	&	1.104	&	0.803	&	0.278	\\
			&		&	$\gamma$	&	0.317	&	0.152	&	0.069	&	0.051	&	0.071	&	0.136	&	0.290	&	--	&	--	\\\\
			&	0.1	&	$\beta_1$	&	1.273	&	0.672	&	0.283	&	0.170	&	0.259	&	0.554	&	1.224	&	1.730	&	0.677	\\
			&		&	$\beta_2$	&	0.791	&	0.370	&	0.159	&	0.110	&	0.197	&	0.562	&	1.116	&	1.359	&	0.639	\\
			&		&	$\gamma$	&	0.436	&	0.206	&	0.088	&	0.057	&	0.085	&	0.159	&	0.328	&	--	&	--	\\	
			\hline
		\end{tabular}
	}
	\resizebox{.85\textwidth}{!}{
	\begin{tabular}{lll|rrrrrrr|rr}\hline
		\multicolumn{12}{c}{$n=100$}\\\hline
		5\% 	&	0	&	$\beta_1$	&	0.012	&	0.012	&	0.014	&	0.021	&	0.042	&	0.096	&	0.394	&	0.046	&	0.078	\\
		&		&	$\beta_2$	&	0.011	&	0.011	&	0.013	&	0.020	&	0.042	&	0.111	&	0.576	&	0.045	&	0.082	\\
		&		&	$\gamma$	&	0.012	&	0.012	&	0.013	&	0.017	&	0.028	&	0.056	&	0.112	&	--	&	--	\\\\
		&	0.05	&	$\beta_1$	&	1.021	&	0.184	&	0.042	&	0.032	&	0.075	&	0.190	&	0.494	&	1.387	&	0.205	\\
		&		&	$\beta_2$	&	0.548	&	0.106	&	0.030	&	0.027	&	0.060	&	0.211	&	0.537	&	1.017	&	0.206	\\
		&		&	$\gamma$	&	0.341	&	0.068	&	0.020	&	0.017	&	0.029	&	0.083	&	0.232	&	--	&	--	\\\\
		&	0.1	&	$\beta_1$	&	1.639	&	0.422	&	0.099	&	0.069	&	0.140	&	0.304	&	0.569	&	2.143	&	0.611	\\
		&		&	$\beta_2$	&	1.026	&	0.202	&	0.053	&	0.039	&	0.079	&	0.188	&	0.537	&	1.713	&	0.588	\\
		&		&	$\gamma$	&	0.533	&	0.130	&	0.032	&	0.022	&	0.033	&	0.108	&	0.341	&	--	&	--	\\
		\hline\\
		10\% 	&	0	&	$\beta_1$	&	0.013	&	0.013	&	0.016	&	0.025	&	0.048	&	0.110	&	0.433	&	0.050	&	0.087	\\
		&		&	$\beta_2$	&	0.013	&	0.014	&	0.017	&	0.027	&	0.051	&	0.186	&	0.604	&	0.046	&	0.083	\\
		&		&	$\gamma$	&	0.017	&	0.019	&	0.022	&	0.033	&	0.050	&	0.085	&	0.138	&	--	&	--	\\\\
		&	0.05	&	$\beta_1$	&	0.992	&	0.188	&	0.060	&	0.037	&	0.079	&	0.183	&	0.473	&	1.381	&	0.198	\\
		&		&	$\beta_2$	&	0.565	&	0.096	&	0.038	&	0.028	&	0.066	&	0.275	&	0.456	&	1.028	&	0.195	\\
		&		&	$\gamma$	&	0.366	&	0.087	&	0.040	&	0.034	&	0.049	&	0.091	&	0.189	&	--	&	--	\\\\
		&	0.1	&	$\beta_1$	&	1.622	&	0.468	&	0.124	&	0.077	&	0.154	&	0.324	&	0.622	&	2.115	&	0.589	\\
		&		&	$\beta_2$	&	1.047	&	0.230	&	0.071	&	0.038	&	0.083	&	0.201	&	0.736	&	1.711	&	0.577	\\
		&		&	$\gamma$	&	0.567	&	0.170	&	0.058	&	0.037	&	0.048	&	0.098	&	0.261	&	--	&	--	\\	
		\hline
	\end{tabular}
}
	\label{TAB:MDPDE_MSE_n50}
\end{table}

\begin{table}[h]
	\centering
\caption{Parameter estimates for the Myeloma data}
	\resizebox{0.9\textwidth}{!}{
		\begin{tabular}{l|rrrrrrr|rr}\hline
			& \multicolumn{7}{c|}{Parametric MDPDE with $\alpha$} & \multicolumn{2}{c}{Semi-parametric}\\		
			&	0 (MLE)	&	0.05	&	0.1	&	0.2	&	0.3	&	0.4	&	0.5	&	PLE	&	BRE	\\\hline
			\multicolumn{8}{l}{\underline{Full Data With Outliers}} &&\\
			BUN	&	0.016	&	0.016	&	0.016	&	0.016	&	0.016	&	0.016	&	0.016	&	0.023	&	0.025	\\
			CALC	&	0.137	&	0.155	&	0.173	&	0.207	&	0.234	&	0.253	&	0.265	&	0.165	&	0.298\\	
			HGB	&	$-$0.059	&	$-$0.065	&	$-$0.073	&	$-$0.091	&	$-$0.110	&	$-$0.128	&	$-$0.141	&	$-$0.137	&	$-$0.180	\\
			$\gamma$	&	0.012	&	0.010	&	0.009	&	0.007	&	0.006	&	0.006	&	0.006	&	--	&	--	\\
			\hline
			\multicolumn{8}{l}{\underline{Without 3 outlying observations}} &&\\
			BUN	&	0.016	&	0.016	&	0.016	&	0.015	&	0.015	&	0.015	&	0.015	&	0.027	&	0.025	\\
			CALC	&	0.247	&	0.251	&	0.254	&	0.261	&	0.267	&	0.273	&	0.278	&	0.370	&	0.344	\\
			HGB	&	$-$0.123	&	$-$0.128	&	$-$0.133	&	$-$0.145	&	$-$0.156	&	$-$0.166	&	$-$0.175	&	$-$0.235	&	$-$0.218	\\
			$\gamma$	&	0.009	&	0.009	&	0.009	&	0.009	&	0.008	&	0.008	&	0.008	&	--	&	--	\\
			\hline
		\end{tabular}
	}
	\label{TAB:Data_Myeloma}
\end{table}

\begin{table}[h]
	\centering
	\caption{Parameter estimates for the Breast and Ovarian Cancer data}
	\resizebox{0.9\textwidth}{!}{
		\begin{tabular}{l|rrrrrrr|rr}\hline
			& \multicolumn{7}{c|}{Parametric MDPDE with $\alpha$} & \multicolumn{2}{c}{Semi-parametric}\\		
			&	0 (MLE)	&	0.05	&	0.1	&	0.2	&	0.3	&	0.4	&	0.5	&	PLE	&	BRE	\\\hline
			\multicolumn{8}{l}{\underline{Full Data With Outliers}} &&\\
Type & $-$1.585	&	$-$2.372	&	$-$2.583	&	$-$2.979	&	$-$3.390	&	$-$3.886	&	$-$4.894	&	$-$1.570	&	$-$1.770	\\
$\gamma_1$	&	0.136	&	0.131 &	0.060	&	0.011	&	0.002	&	1.84e$^{-4}$	&	1.14e$^{-5}$	&	--	&	--	\\
$\gamma_2$ & 1.361	&	0.087	&	0.175	&	0.336	&	0.467	&	0.572	&	0.662	&	--	&	--	\\
			\hline
			\multicolumn{8}{l}{\underline{Without 19 outlying observations}} &&\\
Type 	& $-$1.762	&	$-$2.389	&	$-$2.608	&	$-$3.048	&	$-$3.507	&	$-$4.047	&	$-$4.645	&	$-$1.840	&	$-$1.800	\\
$\gamma_1$ &  1.6e$^{-5}$	&	0.001	&	0.060	&	0.011	&	0.002	& 1.73e$^{-4}$	&	1.1e$^{-5}$	&	--	&	--	\\
$\gamma_2$ & 1.497	&	0.879	&	0.176	&	0.345	&	0.485	&	0.597	&	0.686	&	--	&	--	\\
			\hline
		\end{tabular}
	}
	\label{TAB:Data_BRCAOV}
\end{table}

\end{document}